\documentclass[11pt]{article}
\usepackage[letterpaper,margin=1in]{geometry}

\usepackage[utf8]{inputenc}

\usepackage[titles,subfigure]{tocloft}
\usepackage{amsmath,amssymb,amsthm,amsfonts}
\usepackage{mathtools}
\usepackage{graphicx} 
\usepackage{float}
\usepackage{subfigure} 
\usepackage{arydshln}
\usepackage{multirow}
\usepackage{authblk,cite}
\usepackage{bm} 
\usepackage{mathrsfs}   
\usepackage{setspace}
\usepackage{url}
\usepackage{hyperref}
\hypersetup{colorlinks,linkcolor=blue,urlcolor=blue,citecolor=blue}
\usepackage{url}
\usepackage{xcolor}
\usepackage[title]{appendix}
\usepackage{qcircuit}

\newcommand{\Qd}{Q^{\mathrm{DQC1}}}
\newcommand{\Qb}{Q^{\mathrm{BQP}}}

\usepackage{tikz}  %graphics
\usepackage{tikz-cd}  %commutative diagram
\usetikzlibrary{positioning}
\usepackage{tikz-qtree}
\usepackage{mathdots}
\usepackage{authblk}

\usepackage{qcircuit}
\newcommand{\ket}[1]{| #1 \rangle}
\newcommand{\bra}[1]{\langle #1|}

\usepackage{titletoc}

\DeclareMathOperator{\tr}{Tr}

\DeclareMathOperator{\sgn}{sgn}

\newcommand{\F}{\mathbb{F}}

\def\dego{\widetilde{\deg}^{\mathrm{DQC1}}}

\allowdisplaybreaks

\newtheorem{theorem}{Theorem}[section]
\newtheorem{proposition}[theorem]{Proposition}
\newtheorem{corollary}[theorem]{Corollary}
\newtheorem{lemma}[theorem]{Lemma}

\theoremstyle{definition}

\newtheorem{definition}[theorem]{Definition}
\newtheorem{remark}[theorem]{Remark}

\numberwithin{equation}{section}

\newtheorem*{thm*}{Theorem}
\newtheorem*{lem*}{Lemma}
\newtheorem*{prop*}{Proposition}
\newtheorem*{defn*}{Definition}
\newtheorem*{prob*}{Problem}
\newtheorem*{ques*}{Question}

\newcommand{\be}{\begin{equation}}
\newcommand{\ee}{\end{equation}}
\newcommand{\bes}{\begin{equation*}}
\newcommand{\ees}{\end{equation*}}
\newcommand{\bea}{\begin{eqnarray}}
\newcommand{\eea}{\end{eqnarray}}
\newcommand{\beas}{\begin{eqnarray*}}
\newcommand{\eeas}{\end{eqnarray*}}
\newcommand{\bal}{\begin{aligned}}
\newcommand{\eal}{\end{aligned}}

\usepackage[title]{appendix}

\title{ A sensitivity-constrained polynomial method for DQC1 query complexity }

\author{Changpeng Shao\thanks{\texttt{changpeng.shao@amss.ac.cn}}}

\affil{\small{SKLMS, Academy of Mathematics and Systems Science, Chinese Academy of Sciences, Beijing, China}}

\date{\today}
\begin{document}

\maketitle

\begin{abstract}
The polynomial method is one of the fundamental techniques for proving lower bounds on quantum query complexity, particularly in the BQP query model. In this work, we develop an analogue of the polynomial method for the DQC1 query model. We introduce the notion of {\em sensitivity approximate degree} for Boolean functions and show that it provides a lower bound on DQC1 query complexity. We establish several basic structural properties of this new measure.
As a first result, we show that every nonconstant total Boolean function on \(N\) input variables has sensitivity approximate degree \(\Omega(N)\), and hence requires \(\Omega(N)\) queries in the DQC1 model. Thus, unlike in the BQP query model, DQC1 admits no asymptotic query speedup for total Boolean functions.
For symmetric partial Boolean functions, we give a complete characterization of the sensitivity approximate degree in terms of the separation between oppositely labeled Hamming weight layers. This yields a tight quadratic separation between sensitivity approximate degree and standard approximate degree, and also leads to tight quadratic separations between DQC1 and BQP query complexities.  
We further establish quantitative average-case classical simulation theorems for DQC1 algorithms. 
%Specifically, for any $T$-query DQC1 algorithm and any product or exchangeable input distribution $\mu$, there exists a deterministic, nonadaptive classical algorithm using $O(T^2\varepsilon^{-2}\log(1/\delta))$ queries that approximates the quantum acceptance probability to additive error at most $\varepsilon$ on at least a $1-\delta$ fraction of inputs drawn from $\mu$.
Finally, we apply our framework to Fourier analysis, obtaining a simple proof of an $\Omega(N^{1/4})$ DQC1 query lower bound for the 2-Forrelation problem.

% ~\\

% {\bf Keywords.} DQC1, quantum query complexity, polynomial method, average-case classical simulation

% {\bf MSC codes.} Primary: 68Q12; Secondary: 68Q17, 68Q25, 41A10
\end{abstract}

% \begingroup
% \let\bfseries\mdseries
% \offinterlineskip
% \setcounter{tocdepth}{1}
% \tableofcontents
% \endgroup

% \begingroup
% \setstretch{0.01}
% \setcounter{tocdepth}{1}
% \tableofcontents
% \endgroup

\section{Introduction}

Quantum query complexity provides a clean framework for understanding the computational power and limitations of quantum algorithms. 
%By counting only the number of accesses to an input oracle, the query model isolates the information-theoretic difficulty of a problem from the cost of implementing the remaining computation.  
It has played a central role in the development of quantum algorithms and lower-bound techniques, from early examples such as the Deutsch--Jozsa, Bernstein--Vazirani, and Simon algorithms to Grover search and element distinctness, as well as foundational lower-bound methods such as the polynomial and adversary methods~\cite{deutsch1992rapid, bernstein1993quantum, simon1997power, grover1996fast, ambainis2007quantum, beals2001quantum, Ambainis2002, HoyerLeeSpalek2007}. More broadly, query complexity has become a useful testing ground for understanding the structure and limitations of quantum speedups~\cite{AaronsonAmbainis2014,buhrman2002complexity}.

One of the fundamental tools for studying quantum query complexity is the polynomial method introduced in~\cite{beals2001quantum}. The key observation is that the acceptance probability of a $T$-query quantum algorithm can be represented by a real multilinear polynomial of degree at most $2T$ in the input variables. Consequently, lower bounds on the degree of polynomials approximating a Boolean function translate directly into quantum query lower bounds. This connection led naturally to the study of approximate degree~\cite{NisanSzegedy1994, Paturi1992}, which has become one of the standard tools for proving lower bounds in quantum query complexity \cite{childs2017lecture, de2019quantum}. Thus, the polynomial method replaces the analysis of a quantum computation with a problem in real polynomial approximation, making available techniques from approximation theory, Boolean function analysis, and Fourier analysis~\cite{BunThaler2022, ODonnell2014}.

While the standard BQP query model provides a powerful framework for studying quantum speedups, it is also important to understand which quantum resources are essential for achieving them. One way to investigate this question is to study computational models with restrictions on resources such as circuit depth, ancillas, or the purity of the initial quantum state.
In this work, we focus on the one-clean-qubit model, or Deterministic Quantum Computation with One Qubit (DQC1)~\cite{knill1998power}, which is one of the simplest restricted models of quantum computation that nevertheless exhibits nontrivial computational power.
In this model, the computation starts with one pure qubit
and $m$ maximally mixed qubits, in the state
\be
\rho_0
=
|0\rangle\langle0|\otimes\frac{I_{2^m}}{2^m}.
\ee
A unitary $U$ is then applied to all $m+1$ qubits, followed by a computational-basis measurement of the clean qubit. Taking outcome $0$ as acceptance, and
writing $Z=Z_{\mathrm{clean}}\otimes I_{2^m}$, the output bias is 
\be
p=\operatorname{Tr}(ZU\rho_0U^\dagger)
=\frac{1}{2^{m+1}} \operatorname{Tr}(ZUZU^\dagger) .
\ee
In the query setting, $U=U_x$ consists of oracle calls
interleaved with input-independent unitaries.

Despite the severe restriction, DQC1 can efficiently perform tasks such as normalized trace estimation, and several natural problems, including approximations of the Jones polynomial and normalized trace of functions of matrices, are complete for the model \cite{knill1998power, shor2008estimating, brandao2008entanglement, cade, edenhofer2025dequantization,  ji2026dqc1}.  DQC1 therefore provides a
useful setting for studying how much computational power can survive in the presence of very limited purity.

% While DQC1 has been extensively studied from the perspective of computational complexity, its query complexity has received comparatively little attention. Existing works in the query setting have mainly focused on specific problems or particular structural properties of DQC1 algorithms~\cite{Shepherd2009, Girish2026Fourier, ji2026dqc1, zhang-shao}. To the best of our knowledge, a systematic general framework for studying DQC1 query complexity is still lacking. Developing such a framework is particularly useful for comparing DQC1 with standard quantum and classical query models, and for identifying which quantum query advantages can survive under the one clean qubit restriction.

While DQC1 has been extensively studied from the perspective of computational complexity, its query complexity has received comparatively little attention.
Existing works in the query setting have mainly focused on specific problems, oracle separations, or structural properties of DQC1 query algorithms
\cite{Shepherd2009,Girish2026Fourier,ji2026dqc1,zhang-shao}.
Despite these advances, a general-purpose framework for studying the query complexity of arbitrary Boolean functions in the DQC1 model remains largely undeveloped. In particular, a polynomial framework analogous to the standard polynomial method for quantum query complexity has not been systematically
developed for the DQC1 query model. Such a framework would provide a direct way to compare DQC1 with standard quantum and classical query models and to identify which quantum query advantages can survive under the one clean qubit restriction.

We address this gap by developing a polynomial method tailored to the DQC1 query model. The key observation here is that, unlike in the standard quantum query model, a maximally mixed query register cannot concentrate its weight on a chosen input coordinate. As a consequence, the output polynomial of a \(T\)-query DQC1 algorithm satisfies not only the usual degree bound, but also a uniform coordinate sensitivity bound. 
More precisely, the output bias of any $T$-query DQC1 algorithm is a real multilinear polynomial $p:\{-1,1\}^N\to[-1,1]$ satisfying
\be
\deg (p)\leq 2T,
\qquad
\|p\|_\infty\leq 1,
\qquad
\max_{i\in[N]}\|\partial_i p\|_\infty
\leq \frac{T}{N},
\ee
where $\partial_i p$ denotes the Boolean derivative in coordinate $i$, and the supremum norms are taken over the entire Boolean cube.
The first two constraints are shared by standard quantum query algorithms and therefore do not, by themselves, reflect the additional restrictions of DQC1. The sensitivity bound, in contrast, limits how strongly the output can respond to changes in any individual input coordinate. 
Thus, a polynomial may have low degree yet be too sensitive to individual coordinates to arise from a low-query DQC1 algorithm. By incorporating this additional constraint into polynomial approximation, our method yields DQC1 query lower bounds that ordinary approximate degree alone cannot establish. 
This motivates the notion of sensitivity approximate degree, formalized in Definition~\ref{def:dqc1-sens-degree}.

We establish a range of structural properties of the sensitivity approximate degree and derive several consequences for DQC1 query complexity. We highlight four representative results below.

\paragraph{Linear query lower bounds for total Boolean functions.}
Our first result shows that every nonconstant total Boolean function on $N$ variables requires $\Omega(N)$ queries in the DQC1 model; see Corollary \ref{cor:total-functions}. Thus, unlike in the standard BQP query model,
DQC1 admits no asymptotic quantum query speedup for total Boolean functions. This provides a striking illustration of the limitations of DQC1: even the dictator function, which has polynomial degree one and can be computed with a single classical query, requires $\Omega(N)$ DQC1 queries. Consequently, any asymptotic quantum query speedup in the DQC1 model must exploit input structure encoded by a promise, motivating the study of partial Boolean functions.
The lower bound follows from a general relation between sensitivity approximate degree and the minimum Hamming distance between oppositely labeled inputs established in Theorem \ref{thm:hamming-distance-lower-bound}.

\paragraph{Characterization of sensitivity approximate degree for symmetric functions.}
%We develop a systematic theory of sensitivity approximate degree, investigating its behavior under restrictions, padding, replication.
%, and block composition.
For symmetric partial Boolean functions, we obtain a complete characterization of the sensitivity approximate degree, up to constant factors, in terms of the minimum separation between oppositely labeled Hamming-weight layers.
Specifically, we show that the DQC1 sensitivity approximate degree of any nonconstant symmetric function $f$ is $\Theta(N/\Delta_H(f))$, where 
$\Delta_H(f):=
\min_{f(x)\neq f(y)}d_H(x,y)$; see Theorem \ref{thm:symmetric-fixed-error}.
This leads to a tight quadratic relationship between sensitivity approximate degree and standard approximate degree; see Corollary \ref{cor:symmetric-quadratic-comparison}. 
We further establish in Theorem \ref{thm:Quadratic query comparison} that
\be
Q^{\mathrm{DQC1}}(f)
=O\bigl(Q^{\mathrm{BQP}}(f)^2\bigr)
\ee
for every symmetric partial Boolean function, and show that this quadratic dependence is tight. Here, $Q^{\mathrm{DQC1}}(f)$ and $Q^{\mathrm{BQP}}(f)$
denote the query complexities of $f$ in the DQC1 and BQP
models, respectively.

% Beyond the symmetric setting, we demonstrate that Hamming-distance geometry alone is insufficient to characterize sensitivity approximate degree. Using linear codes, we construct partial Boolean functions with ordinary approximate degree one and minimum bichromatic Hamming distance $\Omega(N)$, yet sensitivity approximate degree $\Omega(N)$. This reveals an obstruction arising from algebraic constraints on the promise domain that is not captured by bichromatic Hamming distance.
% We also derive Fourier-growth bounds from the coordinate-sensitivity constraint, recovering a nontrivial DQC1 query lower bound for 2-Forrelation through a simple Fourier-analytic argument.

\paragraph{Average-case classical simulation of DQC1 algorithms.}
The recent resolution of the Aaronson Ambainis conjecture \cite{turturean2026proof}
implies that general quantum query algorithms admit average-case
classical simulations with polynomial query overhead under the
uniform input distribution.
For DQC1 algorithms, we establish stronger quantitative guarantees
that apply to broader classes of input distributions.
Our proofs rely only on the coordinate-sensitivity constraint
and standard concentration inequalities, without using
the polynomial degree bound or an influential-variable
argument.
Specifically, we show that the acceptance probability of any
$T$-query DQC1 algorithm can be approximated, with error at most
$\varepsilon$ and failure probability at most $\delta$, by a
deterministic, nonadaptive classical algorithm using
\be
O\left(
\frac{T^2}{\varepsilon^2}\log\frac1\delta
\right)
\ee
queries under arbitrary product or exchangeable input distributions;
see Theorems \ref{thm:dqc1-product-simulation} and
\ref{thm:dqc1-exchangeable-simulation}.
Moreover, under exchangeable distributions, the same simulator
works independently of the input distribution.
We also establish an approximate symmetry property:
the acceptance probability is close, in a distributional sense,
to a Lipschitz function of the normalized Hamming weight;
see Corollary \ref{cor:approximate-symmetry}.
These results rule out superpolynomial average-case DQC1--classical
query separations under these distributions and highlight the
strong structural restrictions imposed by the DQC1 model.

\paragraph{A simple proof of the DQC1 lower bound for 2-Forrelation.}
We give a short Fourier-analytic proof of an $\Omega(N^{1/4})$ DQC1 query lower bound for 2-Forrelation (Corollary \ref{cor:two-forrelation-dqc1}), a problem studied in this model by Girish~\cite{Girish2026Fourier}.\footnote{
The corrected arXiv version (v3) of~\cite{Girish2026Fourier} states a $\widetilde{\Omega}(N^{1/6})$ DQC1 query lower bound for 2-Forrelation. In subsequent email correspondence, Uma Girish informed us of an argument improving the level-two Fourier-growth bound from $O(T^3)$ to $O(T^2)$, which would yield a
$\widetilde{\Omega}(N^{1/4})$ lower bound.
This improvement is not included in the cited version.}
The key step is an elementary $O(T^2)$ bound on the level-two Fourier growth of the output polynomial of any $T$-query DQC1 algorithm, uniformly under restrictions; see Proposition \ref{prop:level-two-fourier-growth}.
This bound follows directly from the coordinate-sensitivity constraint, Bernstein's inequality, and a derivative-counting identity, without requiring further analysis of the underlying matrix structure.
Combining it with the existing reduction from Fourier-growth to Forrelation of Girish, Raz, and Zhan \cite{GirishRazZhan2021} yields the desired query lower bound.
In fact, the argument establishes an $\Omega(N^{1/4})$ lower bound already for sensitivity approximate degree 
(Theorem \ref{thm:two-forrelation-sensitivity}), showing that degree and coordinate sensitivity alone capture the obstruction needed for this application.

\paragraph{Outline of the paper.}
Section~\ref{sec:DQC1 sensitivity approximate degree} introduces the DQC1 query model, the sensitivity-constrained polynomial method together with some basic properties.
Section~\ref{sec:symmetric-functions} gives a tight characterization for symmetric partial functions and a quadratic relation between DQC1 and standard quantum query complexities.
% Section~\ref{sec:linear-code-separation} gives a linear-code separation showing that bichromatic geometry does not characterize the general case.
% Section~\ref{sec:sens-composition-lower} studies composition rules and their sharpness.
Section \ref{sec:dqc1-average-case-simulation} establishes concentration and approximate symmetry properties of DQC1 acceptance probabilities, leading to average-case classical simulation under product and exchangeable input distributions.
Finally, Section~\ref{sec:Fourier analysis} develops Fourier consequences of coordinate sensitivity and applies them to $2$-Forrelation.

\section{DQC1 sensitivity approximate degree}
\label{sec:DQC1 sensitivity approximate degree}

Let $N=2^n$ and $x=(x_1,\ldots,x_N)\in\{1,-1\}^N$. We consider the following standard controlled-phase oracle \cite{de2019quantum,childs2017lecture}:
\be
O_x:  \ket{b} \ket{i} \mapsto x_i^b \ket{b} \ket{i}, \quad b\in \{0,1\}, i\in[N].
\label{eq:oracle}
\ee
In matrix form,
\be
\label{oracle}
O_x = P_0 + \sum_{i=1}^N x_i P_i,
\ee
where
\be
P_0 = \ket{0} \bra{0} \otimes I_N, \qquad P_i = \ket{1} \bra{1} \otimes \ket{i} \bra{i}, \,\, i \in [N].
\ee
They also satisfy $P_i P_j = \delta_{ij} P_i$ and $\sum_{i=0}^N P_i = I_{2N}.$

\begin{definition}
For a $d\times d$ matrix $A$, define its normalized trace as 
\be
\tau(A) := \frac{\tr[A]}{d}.
\ee
\end{definition}

With the above notation, we have 
\be
\label{eq:trace of Pi}
\tau(P_0) = \frac{1}{2}, \qquad \tau(P_i) = \frac{1}{2N}, \quad i\in [N].
\ee

A DQC1 query algorithm acts on one clean qubit and $m\geq n$ maximally mixed qubits, initially prepared in the state
\be
\rho_0
=
|0\rangle\langle0|
\otimes\frac{I_{2^m}}{2^m}.
\ee
The algorithm may use arbitrary input-independent unitaries and any number of additional maximally mixed qubits, but no additional clean qubits. We use the above standard controlled phase oracle \eqref{oracle}, where the control qubit may be the unique clean qubit. Each oracle invocation counts as one query. The output is obtained by measuring the clean qubit at the end of the computation.
More clearly, a $T$-query DQC1 algorithm consists of arbitrary input-independent unitaries $V_0,\ldots,V_T$, interleaved with $T$ applications of the oracle $O_x$. The overall unitary is given by
\be
U_x=V_T O_x V_{T-1}\cdots V_1 O_x V_0.
\ee
Here $O_x$ should be viewed as $O_x\otimes I_{m-n}$.
At the end of the computation, the clean qubit is measured in
the computational basis. Let
$Z=Z_{\mathrm{clean}}\otimes I_{2^m}.$
The output bias of the algorithm is defined as
\be
p(x)=\operatorname{Tr}\left(
ZU_x\rho_0 U_x^\dagger
\right).
\ee
Since $\rho_0={(I+Z)}/{2^{m+1}}$
and $\operatorname{Tr}(Z)=0$, we obtain
\be 
\label{eq:dqc1-output-bias}
p(x)=\tau\left(ZU_xZU_x^\dagger\right).
\ee

\begin{remark}
We focus on the DQC1 model throughout this paper. Indeed, extending our polynomial method to the DQC$k$ model, with $k$ clean qubits, introduces at most an additional factor of $2^k$ in the coordinate-sensitivity bound in \eqref{eq:dqc1-degree-sensitivity} and \eqref{eq:dqc1-sens-degree4} below. In particular, when $k=O(\log\log N)$, a regime in which DQC$k$ and DQC1 are computationally equivalent \cite{shor2008estimating}, this amounts to at most a polylogarithmic factor in $N$. We therefore restrict our attention to DQC1 to highlight the essential features of the polynomial method.
\end{remark}

\begin{definition}[Bounded-error DQC1 query complexity]
For a partial Boolean function
$f:D \subseteq \{-1,1\}^N \to\{-1,1\}$, we say that a DQC1 algorithm computes
$f$ with error at most $\varepsilon$ if
\be
\Pr[\mathrm{output}=f(x)]\geq 1-\varepsilon,
x\in D.
\ee
Equivalently, its output bias \eqref{eq:dqc1-output-bias} satisfies
\be
|p(x)-f(x)|\leq 2 \varepsilon,
\qquad x\in D.
\ee
We denote by $Q_\varepsilon^{\mathrm{DQC1}}(f)$ the
$\varepsilon$-error DQC1 query complexity of $f$,
defined as the minimum number of oracle queries
required by a DQC1 algorithm that computes $f$
with error probability at most $\varepsilon$.
For the standard bounded-error setting, we write
$Q^{\mathrm{DQC1}}(f)
:=Q_{1/3}^{\mathrm{DQC1}}(f).$
\end{definition}

\begin{theorem}
\label{thm:dqc1-derivative}
Let $p:\{-1,1\}^N\to[-1,1]$ be the output bias
of a $T$-query DQC1 algorithm. Then $p(x)$ is a real
multilinear polynomial satisfying
\begin{equation}
\label{eq:dqc1-degree-sensitivity}
\deg (p)\leq 2T,
\qquad
\|p\|_\infty\leq 1,
\qquad
\max_{i\in[N]}\|\partial_i p\|_\infty
\leq \frac{T}{N},
\end{equation}
where
\be
\partial_i p(x)
=
\frac{
p(x^{i\leftarrow +1})
-
p(x^{i\leftarrow -1})
}{2}.
\ee
Here, all supremum norms are taken over the entire
Boolean cube $\{-1,1\}^N$, and \(x^{i\leftarrow \pm1}\) denotes the input obtained from \(x\) by setting its \(i\)-th coordinate to \(\pm1\) while leaving all other coordinates unchanged.
\end{theorem}

\begin{proof}
Similar to the standard polynomial method \cite{beals2001quantum}, the first two claims in \eqref{eq:dqc1-degree-sensitivity} follow directly.  
It remains to prove the coordinate sensitivity
bound. Fix $i\in[N]$ and all input coordinates
other than $x_i$. Write $x^+=x^{i\leftarrow+1}, x^-=x^{i\leftarrow-1},$
and let $O_+=O_{x^+}$ and $O_-=O_{x^-}$.
By the definition of the oracle,
$O_+-O_-=2P_i.$
Consider the difference
\[
p(x^+)-p(x^-)
=
\tau\left(ZU_{x^+}ZU_{x^+}^\dagger\right)
-
\tau\left(ZU_{x^-}ZU_{x^-}^\dagger\right).
\]
Applying the telescoping identity to the
$2T$ oracle occurrences in the two trace
expressions, we obtain a sum of $2T$ terms,
each of the form $2\tau(AP_iB)$,
where $A$ and $B$ are products of unitary
operators and hence satisfy
$\|A\|,\|B\|\leq 1$.
By cyclicity of the trace and the trace-norm
inequality,
\[
|\tau(AP_iB)|
=|\tau(P_iBA)|
\leq \tau(P_i)\|BA\|
\leq \tau(P_i)
=\frac{1}{2N}.
\]
Consequently, by \eqref{eq:trace of Pi},
\[
|p(x^+)-p(x^-)|
\leq
2T\cdot 2\tau(P_i)
=
\frac{2T}{N}.
\]
Dividing by 2 gives
$|\partial_i p(x)|
\leq {T}/{N}.$
Since the bound holds uniformly over all
choices of the remaining input coordinates, we have
$\|\partial_i p\|_\infty
\leq {T}/{N}$.
Taking the maximum over $i\in[N]$ completes
the proof.
\end{proof}

With the above result, we define the following concept of approximate degree in DQC1.

\begin{definition}[DQC1 sensitivity approximate degree]
\label{def:dqc1-sens-degree}
Let $f:D\to\{-1,1\}$ be a Boolean function,
where $D\subseteq\{-1,1\}^N$, and let
$0\leq\varepsilon<1$.
The \emph{DQC1 sensitivity approximate degree}
of $f$, denoted by $\dego_\varepsilon(f)$,
is the minimum integer $d\geq 0$ for which there
exists a real multilinear polynomial
$p:\{-1,1\}^N\to\mathbb{R}$ satisfying
\bea
\deg p \leq d,  \label{eq:dqc1-sens-degree1} \\
|p(x)-f(x)| \leq \varepsilon,
&& \forall x\in D, \label{eq:dqc1-sens-degree2} \\
\|p\|_\infty \leq 1,&&  \label{eq:dqc1-sens-degree3} \\
\max_{i\in[N]}\|\partial_i p\|_\infty
\leq  {d}/{2N}. && \label{eq:dqc1-sens-degree4}
\eea
When $\varepsilon=1/3$, we simply write $\dego(f)$.
\end{definition}

As a direct corollary of Theorem \ref{thm:dqc1-derivative}, we have the following lower bound of DQC1 query complexity.

\begin{corollary}
\label{cor:dqc1-sens-lower-bound}
Let $f:D\to\{-1,1\}$ be a Boolean function,
where $D\subseteq\{-1,1\}^N$.
For any $0\leq \varepsilon <1/2$, we have
\be
Q_\varepsilon^{\mathrm{DQC1}}(f) \geq  
\frac{1}{2} \dego_{2\varepsilon} (f).
\ee
\end{corollary}

Note that the standard approximate degree, denoted by 
\(\widetilde{\deg}_\varepsilon(f)\), which underlies the polynomial method \cite{beals2001quantum} for BQP query complexity, is defined using only the first three conditions \eqref{eq:dqc1-sens-degree1}--\eqref{eq:dqc1-sens-degree3}. The additional coordinate sensitivity constraint \eqref{eq:dqc1-sens-degree4} reflects an intrinsic limitation of the DQC1 query model.

The coordinate derivative bound in Theorem \ref{thm:dqc1-derivative} has a simple interpretation in terms of information dilution in DQC1. The quantity \(\partial_i p\) measures the response of the output bias to changing a single oracle coordinate. In a standard quantum query algorithm, the query register may be concentrated entirely on a chosen coordinate, so this response can be of constant order after a single query. In DQC1, by contrast, the query register is maximally mixed. A single oracle coordinate therefore occupies only an \(O(1/N)\) fraction of the available query space, and unitary evolution cannot concentrate the mixed-state spectrum onto that coordinate. Consequently, each query can change the output bias by at most \(O(1/N)\) in response to a single-coordinate change, leading after \(T\) queries to the bound \(\|\partial_i p\|_\infty\le T/N\). Thus the derivative constraint captures a basic limitation on the local information that a DQC1 computation can extract from an individual oracle coordinate.

\begin{remark}
The sensitivity approximate degree admits an equivalent formulation.
For a real multilinear polynomial \(p\), define
\be
\label{gamma}
\Gamma_N(p)
:=
\max\left\{
\deg p, \,\,\,\,
2N\max_{i\in[N]}\|\partial_i p\|_\infty
\right\}.
\ee
The first term measures the usual algebraic complexity of \(p\), while the second measures its largest response to an individual input coordinate, rescaled by the dimension \(N\). Theorem \ref{thm:dqc1-derivative} therefore states that the output bias \(p\) of every \(T\)-query DQC1 algorithm satisfies
$\Gamma_N(p)\le 2T.$
Thus \(\Gamma_N\) may be viewed as a polynomial relaxation of DQC1 query complexity: besides low degree, a polynomial arising from a DQC1 computation must also have uniformly limited local sensitivity.
With this notation, the DQC1 sensitivity approximate degree can equivalently be written as
\be
\label{another def}
\dego_\varepsilon(f)
=
\min_{p}
\left\lceil \Gamma_N(p)\right\rceil,
\ee
where the minimum is over all real multilinear polynomials \(p\) satisfying \eqref{eq:dqc1-sens-degree2}, \eqref{eq:dqc1-sens-degree3}.

From \eqref{gamma} and \eqref{another def}, the parameter \(d\) in the definition of the DQC1 sensitivity approximate degree should be viewed as a joint complexity parameter, rather than as the algebraic degree of the approximating polynomial itself. 
Thus, although every multilinear polynomial on \(\{-1,1\}^N\) has algebraic degree at most \(N\), the DQC1 sensitivity approximate degree may be larger than \(N\), because the sensitivity term can be the dominant contribution.
For example, for the dictator function \(f(x)=x_i\), the exact representing polynomial \(p(x)=x_i\) has algebraic degree one, but \(\|\partial_i p\|_\infty=1\). Hence the sensitivity constraint \eqref{eq:dqc1-sens-degree4} requires \(d\ge 2N\). This illustrates that a function may be algebraically very simple while still having large DQC1 sensitivity approximate degree due to its strong dependence on an individual oracle coordinate.
\end{remark}

% \section{Elementary structural properties}
% \label{sec:Elementary structural properties}

% In this section, we establish several basic structural properties of the DQC1 sensitivity approximate degree. These include logarithmic error amplification (Theorem~\ref{thm:sens-error-amplification}); a Hamming distance lower bound (Theorem~\ref{thm:hamming-distance-lower-bound}), which in particular implies linear DQC1 query complexity for every nonconstant total Boolean function (Corollary~\ref{cor:total-functions}); restriction monotonicity (Proposition~\ref{prop:restriction-monotonicity}) and a padding law (Proposition~\ref{prop:padding-law}).
%; and the invariance property under input replication (Proposition~\ref{prop:replication-invariance}).

We next establish several basic structural properties of the DQC1 sensitivity approximate degree.
We first show that the sensitivity constrained approximate degree admits logarithmic error amplification. 
An analogous error reduction result for standard approximate degree was established in \cite[Lemma 1]{buhrman2005robust}, see also \cite[Fact A.3]{tal2014shrinkage}. Our proof follows a similar amplification argument, with the additional observation that the amplification polynomial has a Lipschitz constant bounded by its degree, thereby preserving the coordinate sensitivity constraint under composition.

\begin{theorem}%[Logarithmic error amplification]
\label{thm:sens-error-amplification}
Let $f: D\to\{-1,1\}$ be a Boolean
function with $D\subseteq\{-1,1\}^N$, and let $0\le\delta<1$ and $0<\varepsilon<1$.
Then we have
\begin{equation}
\label{eq:sens-amp-quantitative}
\dego_{\varepsilon} (f)
\le
\left \lceil \frac{2}{(1-\delta)^2}
\log\frac{2}{\varepsilon} \right\rceil
\cdot 
\dego_{\delta} (f).
\end{equation}
In particular, when $\delta=1/3$, we obtain
\be
\dego_{\varepsilon}(f)
\le
\left \lceil \frac{9}{2}
\log\frac{2}{\varepsilon} \right\rceil \cdot 
\dego (f).
\ee
\end{theorem}

\begin{proof}
Let $d=\dego_\delta(f)$, and let $p$ be an
admissible polynomial satisfying \eqref{eq:dqc1-sens-degree1}--\eqref{eq:dqc1-sens-degree4}. 
For an odd integer $r$, let $M_r$ denote the majority
amplification polynomial over $[-1,1]$, defined by $M_r(z)=\mathbb{E}[\operatorname{Maj}(Y_1,\ldots,Y_r)],$
where $Y_1,\ldots,Y_r\in\{1,-1\}$ are independent
random variables with $\mathbb{E}[Y_j]=z$.
As in the standard majority amplification argument
\cite[Fact A.3]{tal2014shrinkage}, we have
$\deg M_r\leq r$ and
$|M_r(z)|\leq 1$.
In addition, by the Hoeffding inequality,
$|M_r(z)-\operatorname{sgn}(z)|
\leq 2e^{-rz^2/2}$ for all $z\in[-1,1]\setminus\{0\}.$
Consequently, choosing $r$ to be the smallest odd
positive integer satisfying
$2e^{-r(1-\delta)^2/2}\leq\varepsilon,$
i.e., $r=\lceil \frac{2}{(1-\delta)^2}
\log\frac{2}{\varepsilon} \rceil$.\footnote{If necessary, $r$ is increased by one to ensure that it is odd.}
Moreover, $\|M_r'\|_{L^\infty([-1,1])}\leq r,$
since $M_r$ is the diagonal restriction of the
multilinear extension of majority, whose individual
partial derivatives are bounded by $1$ on $[-1,1]^r$.

Define $q=M_r\circ p$ and multilinearize it on the
Boolean cube. Then
$\deg q\leq rd, \|q\|_\infty\leq 1.$
Since $f(x)p(x)\geq 1-\delta$ for every $x\in D$, we have
$\sgn(p(x))=f(x)$ and $|p(x)|\geq 1-\delta$.
This implies
$|q(x)-f(x)|
\leq 2e^{-r(1-\delta)^2/2}
\leq\varepsilon.$
Finally, by the mean value theorem and the
Lipschitz bound on $M_r$, we obtain
$\|\partial_i q\|_\infty
\leq \|M_r'\|_{L^\infty([-1,1])}
\|\partial_i p\|_\infty
\leq  {rd}/{2N}.$
Thus $q$ is admissible with degree parameter $rd$.
\end{proof}

\begin{theorem}%[Hamming distance lower bound]
\label{thm:hamming-distance-lower-bound}
Let $f:D\to\{-1,1\}$ be a nonconstant 
Boolean function, where
$D\subseteq\{-1,1\}^N$.
Define the minimum bichromatic Hamming distance by
\be
\Delta_H(f)
:=
\min_{\substack{x,y\in D\\f(x)\neq f(y)}}
d_H(x,y),
\label{eq:minimum bichromatic Hamming distance}
\ee
where $d_H(x,y)$ denotes the Hamming distance
between $x$ and $y$.
Then, for any $0\leq\varepsilon<1$,
\begin{equation}
\label{eq:hamming-distance-lower-bound}
\dego_\varepsilon (f)
\geq
2(1-\varepsilon)\frac{N}{\Delta_H(f)}.
\end{equation}
\end{theorem}

\begin{proof}
Let $d=\dego_{\varepsilon} (f)$,
and let $p$ be a corresponding admissible
polynomial. By definition, $|p(x)-f(x)|\leq\varepsilon$ 
for all $x\in D$,
and $\max_{i\in[N]}\|\partial_i p\|_\infty
\leq {d}/{2N}.$
Choose $x,y\in D$ such that
$f(x)\neq f(y)$ and $d_H(x,y)=\Delta_H(f).$
Since $f(x),f(y)\in\{-1,1\}$,
the approximation condition implies $|p(x)-p(y)|
\geq 2(1-\varepsilon).$

Let $r=d_H(x,y)$, and consider a shortest
Hamming path
$x=z^{(0)},z^{(1)},\ldots,z^{(r)}=y,$
where consecutive vertices differ in exactly
one coordinate.
For each $j\in[r]$, let $i_j$ be the coordinate
on which $z^{(j-1)}$ and $z^{(j)}$ differ.
By the definition of the Boolean derivative,
\[
|p(z^{(j)})-p(z^{(j-1)})|
=2|\partial_{i_j}p(z^{(j-1)})|
\leq\frac{d}{N}.
\]
Applying the triangle inequality along the path,
we obtain
\[
|p(x)-p(y)|
\leq
\sum_{j=1}^{r}
|p(z^{(j)})-p(z^{(j-1)})|
\leq\frac{dr}{N}
=\frac{d\Delta_H(f)}{N}.
\]
Putting it all together, we obtain $2(1-\varepsilon)
\leq {d\Delta_H(f)}/{N}$, which proves the claim.
\end{proof}

The intuition behind the Hamming-distance lower bound is that a DQC1 algorithm has limited sensitivity to individual input coordinates, as each coordinate occupies only an $O(1/N)$ fraction of the maximally mixed query register. Consequently, a polynomial with sensitivity degree $d$ can change by at most $d/N$ along each edge of the Boolean cube. Distinguishing two oppositely labelled inputs at Hamming distance $r$ therefore requires $d=\Omega(N/r)$.

\begin{corollary}%[Lower bound for total Boolean functions]
\label{cor:total-functions}
Let $f:\{-1,1\}^N\to\{-1,1\}$ be a
nonconstant total Boolean function. Then, for
every $0\leq\varepsilon<1$,
\be
2 (1-\varepsilon)N
\leq \dego_\varepsilon (f)
\leq 2N.
\label{eq:lower bound of total 1}
\ee
Consequently, for every $0\leq\varepsilon<1/2$,
\be
Q_\varepsilon^{\mathrm{DQC1}}(f)
\geq (1-2\varepsilon)N.
\label{eq:lower bound of total 2}
\ee
\end{corollary}

\begin{proof}
Since $f$ is nonconstant and the Boolean cube is
connected, there exist adjacent inputs $x,y$
such that $f(x)\neq f(y)$. Hence
$\Delta_H(f)=1$, and
Theorem~\ref{thm:hamming-distance-lower-bound}
implies $\dego_\varepsilon(f)
\geq 2(1-\varepsilon)N.$
The claim \eqref{eq:lower bound of total 2} then follows from Corollary \ref{cor:dqc1-sens-lower-bound}. For the upper bound in \eqref{eq:lower bound of total 1}, let $p$ be the unique multilinear polynomial representing $f$.
Then $\deg p\leq N$, $\|p\|_\infty=1$, and $\max_i\|\partial_i p\|_\infty\leq1$. Thus $p$ is admissible with degree parameter $2N$, proving $\dego_\varepsilon(f)\leq2N$.
\end{proof}

This behavior contrasts sharply with the standard
quantum query model. For total Boolean functions,
bounded error quantum and randomized query
complexities are polynomially related \cite{beals2001quantum}, 
yet polynomial quantum speedups
are possible, as exemplified by quantum search
\cite{grover1996fast} and element distinctness
\cite{ambainis2007quantum}.
In contrast, the preceding corollary shows that
every nonconstant total Boolean function requires
$\Omega(N)$ queries in the DQC1 query model.
Since any total Boolean function can be computed
classically using at most $N$ queries, this model
admits no asymptotic query speedup over randomized
algorithms for total Boolean functions. Consequently, to investigate potential quantum query speedups in the DQC1 model, we must turn to partial Boolean functions.

\section{Sensitivity approximate degree of symmetric partial Boolean functions}
\label{sec:symmetric-functions}

We now study symmetric partial Boolean functions,
for which the DQC1 sensitivity approximate degree admits
a much more precise characterization. In particular,
symmetry allows us to reduce the problem to univariate
polynomial approximation and obtain tight bounds
in terms of the minimum separation between
oppositely labeled Hamming-weight layers.
Symmetry also allows us to establish a
quadratic comparison with standard approximate degree,
revealing the extent to which the additional
coordinate-sensitivity constraint can increase
the degree.

Let $f:D\to\{-1,1\}$, where
$D\subseteq\{-1,1\}^N$, be symmetric under
permutations of the input coordinates.
For $x\in\{-1,1\}^N$, define its Hamming weight by
\[
\operatorname{wt}(x)
=
|\{i\in[N]:x_i=-1\}|.
\]
Then there exist a set
$K\subseteq\{0,1,\ldots,N\}$ and a function
$F:K\to\{-1,1\}$ such that
\[
x\in D
\quad\Longleftrightarrow\quad
\operatorname{wt}(x)\in K,
\qquad
f(x)=F(\operatorname{wt}(x)).
\]

Our first result extends the standard symmetrization
argument for approximate degree
(see, e.g., \cite{childs2017lecture})
to the sensitivity-constrained setting.
As in the standard case, symmetrization reduces
the multivariate polynomial approximation problem
to a univariate one without increasing the degree.
The additional coordinate-sensitivity constraint
translates exactly into a discrete-slope bound
on the resulting univariate polynomial.
As we shall see, this additional constraint can
increase the approximate degree by a quadratic
factor, and its effect can be characterized
precisely for symmetric partial Boolean functions.

\begin{theorem}%[Symmetric reduction]
\label{thm:symmetric-reduction}
Let $f$ be a symmetric partial Boolean function
as above. Then $\dego_\varepsilon (f)$
is the minimum integer $d\geq0$ for which there
exists a univariate real polynomial $q$ satisfying
\bea
\deg q
\leq d, &&
\label{eq:sym-degree} \\
|q(k)|
\leq1,
&& k=0,1,\ldots,N,
\label{eq:sym-bounded} \\
|q(k)-F(k)|
\leq\varepsilon,
&& k\in K,
\label{eq:sym-approx} \\
|q(k+1)-q(k)|
\leq {d}/{N},
&& k=0,1,\ldots,N-1.
\label{eq:sym-slope}
\eea
\end{theorem}

\begin{proof}
Let $p$ be an admissible polynomial with degree
parameter $d$. Symmetrize it over all permutations
of the input coordinates:
\[
\overline p(x)
=
\frac{1}{N!}
\sum_{\pi\in S_N}
p(x_{\pi(1)},\ldots,x_{\pi(N)}).
\]
Since both $D$ and $f$ are symmetric,
$\overline p$ still $\varepsilon$-approximates
$f$ on $D$. Moreover, symmetrization does not
increase the polynomial degree, the uniform norm,
or the maximum coordinate sensitivity. Hence
\[
\deg\overline p\leq d,
\qquad
\|\overline p\|_\infty\leq1,
\qquad
\max_i\|\partial_i\overline p\|_\infty
\leq\frac{d}{2N}.
\]

As $\overline p$ is symmetric, there exists
a univariate polynomial $q$ of degree at most
$d$ such that $\overline p(x)=q(\operatorname{wt}(x)).$
This immediately gives
\eqref{eq:sym-degree}--\eqref{eq:sym-approx}.
To obtain the slope condition, fix a coordinate
$i$ and suppose that the remaining $N-1$
coordinates contain exactly $k$ entries equal
to $-1$. Setting $x_i=+1$ and $x_i=-1$ gives
inputs of Hamming weights $k$ and $k+1$,
respectively. Therefore,
$|\partial_i\overline p(x)|
=
|q(k+1)-q(k)|/2,$
which implies \eqref{eq:sym-slope}.
Conversely, suppose that $q$ satisfies
\eqref{eq:sym-degree}--\eqref{eq:sym-slope}.
Define
\[
p(x)
=
q\left(
\frac{N-\sum_{i=1}^N x_i}{2}
\right),
\]
and take its multilinearization on the Boolean
cube. The resulting polynomial has degree at
most $d$ and satisfies the boundedness and
approximation conditions. Moreover, for every
coordinate $i$,
\[
|\partial_i p(x)|
=
\frac12|q(k+1)-q(k)|
\leq\frac{d}{2N},
\]
where $k$ is the Hamming weight of the remaining
$N-1$ coordinates. Thus $p$ is admissible
with degree parameter $d$.
\end{proof}

For a nonconstant symmetric partial Boolean
function, define the minimum separation between
oppositely labelled Hamming-weight layers by
\be
\Delta(F)
:=
\min_{\substack{a,b\in K\\F(a)\neq F(b)}}
|a-b|.
\ee
For symmetric $f$, this distance is exactly the minimum bichromatic Hamming distance $\Delta_H(f)$ defined in \eqref{eq:minimum bichromatic Hamming distance}. The following theorem shows that, for fixed
approximation error, this single parameter
characterizes the sensitivity approximate degree
up to constant factors.

\begin{theorem}
\label{thm:symmetric-fixed-error}
Let $f:D\to\{-1,1\}$ be a nonconstant symmetric
partial Boolean function on $N$ variables.
For every fixed $0<\varepsilon<1$, we have
\be
\dego_\varepsilon (f)
=
\Theta_\varepsilon\left(
\frac{N}{\Delta(F)}
\right).
\ee
\end{theorem}

\begin{proof}
We first prove the lower bound. Since $f$ is
symmetric, the minimum Hamming distance between
oppositely labelled promise inputs is
$\Delta_H(f)=\Delta(F).$
Theorem~\ref{thm:hamming-distance-lower-bound}
therefore gives
\[
\dego_\varepsilon (f)
\geq
2 (1-\varepsilon)\frac{N}{\Delta(F)}.
\]

For the upper bound, 
we shall construct a function $q$ satisfying all the conditions of
Theorem~\ref{thm:symmetric-reduction} below. 
Consider the rescaled grid
$z_k=-1+\frac{2k}{N} $ for $ k=0,1,\ldots,N.$
On these promise points, define
$h(z_k)
=
 (1-\frac{\varepsilon}{2} )F(k)$ for $k\in K,$
and extend $h$ to the entire interval $[-1,1]$ by connecting its values at consecutive promise points with straight lines, and keeping it constant to the left of the smallest promise point and to the right of the largest one.
By construction, $|h(z)|\leq 1-\varepsilon/2$ for all $z\in[-1,1]$. Moreover, $h$ has slope zero between consecutive promise points with the same function value. When their function values differ, their distance is at least $2\Delta(F)/N$, while the difference in their function values is $2(1-\varepsilon/2)$. Hence,
\[
\operatorname{Lip}(h)
\leq \left(1-\frac{\varepsilon}{2}\right)
\frac{N}{\Delta(F)}.
\]
By the Lipschitz-preserving Jackson approximation
theorem
\cite[Theorem~14]{KordaMagronRiosZertuche2025},
there exists a polynomial $P$ of degree
$m=O_\varepsilon ( {N}/{\Delta(F)} )$,
such that
\[
\|P-h\|_{L^\infty([-1,1])}
\leq\frac{\varepsilon}{2},
\qquad
\operatorname{Lip}(P)
=
O_\varepsilon\left(
\frac{N}{\Delta(F)}
\right),
\]
Denote $d
:=
\left\lceil
\max\left\{
m,\,
2\operatorname{Lip}(P)
\right\}
\right\rceil.$
Now define the polynomial on the Hamming-weight
grid by
\[
q(k)=P\left(-1+\frac{2k}{N}\right).
\]
Then $\deg q\leq d$. Using triangle inequality, $|q(k)|\leq1$ for $k=0,\ldots,N,$
and, 
$|q(k)-F(k)|
\leq\varepsilon$ for $k\in K$.
Moreover,
\[
|q(k+1)-q(k)|
\leq
\operatorname{Lip}(P)\frac{2}{N}
\leq\frac{d}{N}.
\]
Thus $q$ satisfies all the conditions of
Theorem~\ref{thm:symmetric-reduction}.
Therefore,
\[
\dego_\varepsilon (f)
\leq
d
=
O_\varepsilon\left(
\frac{N}{\Delta(F)}
\right),
\]
which proves the claim.
\end{proof}

For fixed bounded error, Theorem \ref{thm:symmetric-fixed-error} shows that the sensitivity approximate degree of a symmetric partial Boolean function depends, up to constant factors, only on the minimum separation between oppositely labelled Hamming-weight layers, regardless of their locations. This contrasts with standard approximate degree, which can depend strongly on the positions of these layers. For example, consider two-weight promise problems supported on the layers $\{0,1\}$ and $\{N/2,N/2+1\}$, respectively, with opposite function values on each pair (assuming $N$ is even). Both have $\Delta(F)=1$ and hence sensitivity approximate degree $\Theta_\varepsilon(N)$. However, their standard approximate degrees are $\Theta_\varepsilon(\sqrt{N})$ and $\Theta_\varepsilon(N)$, respectively~\cite{NayakWu1999}. Thus, unlike standard approximate degree, the sensitivity approximate degree is insensitive, up to constant factors, to the locations of the oppositely labelled layers.

The preceding examples illustrate that the sensitivity approximate degree can be substantially larger than the standard approximate degree. Nevertheless, for symmetric partial Boolean functions, the gap between these two measures is at most quadratic, and this quadratic dependence is tight, as shown in the following corollary.

\begin{corollary}%[Quadratic comparison]
\label{cor:symmetric-quadratic-comparison}
Let $f:D\to\{-1,1\}$ be a nonconstant symmetric
partial Boolean function on $N$ variables.
For every fixed $0<\varepsilon<1$,
\be
\widetilde{\deg}_\varepsilon(f)
\leq
\dego_\varepsilon (f)
\leq
O_\varepsilon\left(
\widetilde{\deg}_\varepsilon(f)^2
\right).
\ee
Moreover, the quadratic dependence is tight.
\end{corollary}

\begin{proof}
The first inequality follows directly from the
definition of the DQC1 sensitivity approximate degree.
For the upper bound, set
$d=\widetilde{\deg}_\varepsilon(f).$
Choose two oppositely labelled Hamming-weight layers
\(u,v\in K\) satisfying
$|u-v|=\Delta(F).$
Let \(f_{u,v}\) denote the restriction of \(f\) to
these two layers. Since restricting the promise domain
cannot increase the standard approximate degree,
$d
=
\widetilde{\deg}_\varepsilon(f)
\geq
\widetilde{\deg}_\varepsilon(f_{u,v}).$
The standard approximate-degree lower bound for
two-weight symmetric promises
\cite{NayakWu1999}
implies
$\widetilde{\deg}_\varepsilon(f_{u,v})
=
\Omega_\varepsilon(
\sqrt{{N}/{|u-v|}}).$
Since \(|u-v|=\Delta(F)\), we obtain
$d
=
\Omega_\varepsilon (
\sqrt{{N}/{\Delta(F)}}
),$
and hence ${N}/{\Delta(F)} = O_\varepsilon(d^2)$.
Combining this with
Theorem~\ref{thm:symmetric-fixed-error}, 
we obtain the upper bound.
The tightness follows directly from the OR function in $N$ variables, for which $\widetilde{\deg}_\varepsilon(f) =\Theta(\sqrt{N})$ while 
$\dego_\varepsilon (f)  = \Theta(N)$.
\end{proof}

The quadratic comparison in Corollary \ref{cor:symmetric-quadratic-comparison} shows that, for symmetric
partial Boolean functions, the sensitivity approximate degree
is at most quadratically larger than standard approximate
degree. However, this does not directly imply a corresponding
quadratic relation between DQC1 and BQP query complexities,
since the sensitivity approximate degree provides only a
lower bound on DQC1 query complexity. Establishing such a
relation would require an additional algorithmic upper bound.

%From \cite[Theorem 4.10]{beals2001quantum}, for symmetric total Boolean function, we have $ Q_\varepsilon^{\rm BQP}(f)=\Omega(\sqrt{N})$. Combining Corollary \ref{cor:total-functions}, we have $Q_\varepsilon^{\rm DQC1}(f) = O(N) \leq O(Q_\varepsilon^{\rm BQP}(f)^2)$ holds for all symmetric total Boolean functions. Next, we mainly focus on symmetric partial Boolean functions.

Recall that any symmetric partial function $f$ is determined by a function $F$ defined over a set $K\subset \{0,1,\ldots,N\}$.
For nonconstant $F$, define
\begin{equation}
\Lambda(F)
:=
\max_{\substack{a,b\in K,\ a<b\\F(a)\ne F(b)}}
\,\,
\frac{b(N-a)}{(b-a)^2}.
\label{eq:Lambda}
\end{equation}
%Notice that $\Lambda(F)\ge1$, since $b\ge b-a$ and $N-a\ge b-a$.

%We first use Nayak-Wu's result~\cite{NayakWu1999} to prove a lower bound on $\Qb_\varepsilon(f)$.

\begin{lemma}[Theorem 1.14 of \cite{podder2025fine}]
\label{lem:adversary}
For every fixed $0<\varepsilon<1/2$, we have
$\Qb_\varepsilon(f)
= \Theta_\varepsilon(\sqrt{\Lambda(F)})$
for every nonconstant symmetric partial Boolean function $f$.
\end{lemma}

\begin{proposition}
\label{prop:Quadratic query comparison}
Fix $0<\varepsilon<1/2$.
For every nonconstant symmetric partial Boolean function $f$, we have
\begin{equation}
\Qd_\varepsilon(f)
\le
2\left\lceil\frac{4\Lambda(F)}{\varepsilon}\right\rceil.
\label{eq:explicit-upper}
\end{equation}
\end{proposition}

Combining Lemma \ref{lem:adversary} and Proposition \ref{prop:Quadratic query comparison}, we have the following result.

\begin{theorem}[Quadratic query separation]
\label{thm:Quadratic query comparison}
For every fixed $0<\varepsilon<1/2$ and every nonconstant symmetric partial Boolean function $f$, we have
\begin{equation}
\Qd_\varepsilon(f)
\le
O_\varepsilon
\bigl( \Qb_\varepsilon(f)^2\bigr).
\label{eq:quadratic}
\end{equation}
\end{theorem}

The tightness of the above quadratic separation also follows from the OR function on $N$ variables.
To prove~\eqref{eq:explicit-upper}, we first construct a
randomized classical algorithm for computing $f(x)$ and
then implement it in the DQC1 model.

\begin{lemma}
\label{lem:classical-sampling-symmetric}
Let $f:D\subseteq\{-1,1\}^N\to\{-1,1\}$ be a nonconstant
symmetric partial Boolean function, represented by
$F:K\to\{-1,1\}$ through $f(x)=F(\operatorname{wt}(x))$.
For every $0<\varepsilon<1/2$, there exists a nonadaptive
randomized classical algorithm that computes $f$ with
error probability at most $\varepsilon$ on every promised
input using
\be
t=\left\lceil\frac{4\Lambda(F)}{\varepsilon}\right\rceil
\ee
queries. 
%The algorithm samples $t$ input coordinates independently and uniformly from $[N]$, and its output depends only on the sampled bit values.
\end{lemma}

\begin{proof}
On input $x$ of Hamming weight $k\in K$, choose $J_1,\ldots,J_t$ independently and
uniformly from $[N]$, query the corresponding coordinates,
and define
\[
Y_j=\frac{1-x_{J_j}}{2}=\begin{cases}
    1 & x_{J_j}=-1 \\
    0 & x_{J_j}=1
\end{cases},
\qquad
\widehat{k}=\frac{N}{t}\sum_{j=1}^t Y_j.
\]
Choose the smallest $k^*\in K$ minimizing
$|\widehat{k}-k^*|$, and output $F(k^*)$.

We now analyze the correctness and cost of this algorithm. Note that the algorithm need not recover $k$ exactly; it suffices that $F(k^*)=F(k)$.
All probabilities and expectations below are over the
sampled indices.
The variables $Y_1,\ldots,Y_t$ are independent Bernoulli random variables satisfying $ \Pr[Y_j=1]=k/N, \Pr[Y_j=0]=1-k/N$, so
$\mathbb E[\widehat{k}]=k$ and $\operatorname{Var}(\widehat{k})={k(N-k)}/{t}.$

Now define
\[
d_k=
\min_{\substack{\ell\in K\\F(\ell)\neq F(k)}}
|k-\ell|.
\]
If $F(k^*)\neq F(k)$, then $|k-k^*|\ge d_k$.
By the choice of $k^*$, we also have $|\widehat{k}-k^*|\le|\widehat{k}-k|.$
Consequently,
$d_k
\le |k-k^*|
\le |k-\widehat{k}|+|\widehat{k}-k^*|
\le 2|\widehat{k}-k|.$
By Chebyshev's inequality,
\[
\Pr[F(k^*)\neq F(k)]
\le
\Pr\left[|\widehat{k}-k|\ge\frac{d_k}{2}\right]
\le
\frac{4k(N-k)}{t\,d_k^2}.
\]
We now bound this probability in terms of $\Lambda(F)$.
Choose $\ell\in K$ with $F(\ell)\neq F(k)$ and
$|k-\ell|=d_k$, and set
$a=\min\{k,\ell\}$ and $b=\max\{k,\ell\}$.
Then $b-a=d_k$, $F(a)\neq F(b)$, and
$k(N-k)\le b(N-a)$.
By the definition of $\Lambda(F)$,
\[
\frac{k(N-k)}{d_k^2}
\le
\frac{b(N-a)}{(b-a)^2}
\le \Lambda(F).
\]
It follows that
\[
\Pr[F(k^*)\neq F(k)]
\le \frac{4\Lambda(F)}{t}
\le \varepsilon.
\]
The algorithm uses $t$ queries, all chosen independently
of the observed input values, and the error bound holds
for every promised input.
\end{proof}

\begin{lemma}
\label{lem:dqc1-sampling}
Let $g:\{0,1\}^t\to\{0,1\}$ be a fixed Boolean function.
There exists a DQC1 algorithm using $2t$ oracle queries
whose output, on every input $x\in\{-1,1\}^N$, has
exactly the same distribution as
\[
g(Y_1,\ldots,Y_t),
\qquad
Y_j=\frac{1-x_{J_j}}{2},
\]
where $J_1,\ldots,J_t$ are independent and uniform over $[N]$.

% More precisely, the algorithm starts in the state
% \[
% \rho_0
% =
% |0\rangle\langle0|_C\otimes\frac{I_{2^m}}{2^m},
% \qquad
% m=tn+t+2^t,
% \]
% applies a unitary $U_x$ consisting of $2t$ oracle calls
% interleaved with input-independent unitaries, and measures
% the clean output qubit $C$ in the computational basis.
\end{lemma}

\begin{proof}
We first specify the registers and initial state, then
construct the DQC1 circuit $U_x$, and finally verify its
output distribution.

\paragraph{Registers and initial state.}
Besides the clean output qubit, use $t$ address
registers $J_1,\ldots,J_t$, each containing $n$ qubits;
a $t$-qubit mask register $R$; and a table register
$D=(D_s)_{s\in\{0,1\}^t}$ containing $2^t$ qubits.
All these workspace registers are maximally mixed.
Writing $J=J_1\cdots J_t$, the initial state is
\[
\rho_0
=
|0\rangle\langle0|_{\text{clean}}
\otimes\frac{I_J}{N^t}
\otimes\frac{I_R}{2^t}
\otimes\frac{I_D}{2^{2^t}}.
\]
This is precisely a DQC1 initial state with
$m=tn+t+2^t$ mixed qubits.
The address registers satisfy
\[
\frac{I_J}{N^t}
=
\frac{1}{N^t}
\sum_{\mathbf i\in[N]^t}
|\mathbf i\rangle\langle\mathbf i|,
\qquad
\mathbf i=(i_1,\ldots,i_t).
\]
For each address tuple $\mathbf i$, define
\[
y(\mathbf i)
=
\left(
\frac{1-x_{i_1}}{2},\ldots,
\frac{1-x_{i_t}}{2}
\right).
\]
Our goal is to construct $U_x$ so that it leaves the
address registers unchanged and, on each address branch
$\mathbf i$, writes $g(y(\mathbf i))$ to the clean qubit,
independently of the initial values of $R$ and $D$.
The address registers are not measured; the branchwise
description is used only to analyze the circuit.

\paragraph{Construction of the unitary.}
Write $r\in\{0,1\}^t$ and
$d=(d_s)_{s\in\{0,1\}^t}\in\{0,1\}^{2^t}$
for computational-basis values of $R$ and $D$.
All operations below act as the identity on unspecified
registers.

First, define
\[
L_x
=
\prod_{j=1}^t
\left(
H_{R_j}\,O_x^{(R_j,J_j)}\,H_{R_j}
\right),
\]
where $H_{R_j}$ is the Hadamard gate acting on qubit $R_j$, and $O_x^{(R_j,J_j)}$ denotes one application of the input oracle with $R_j$ as its control qubit and $J_j$ as its address register:\footnote{
This is an application of the oracle defined in \eqref{eq:oracle},
with $R_j$ and $J_j$ serving as the control qubit and address register, respectively. If necessary, these registers can be routed to the oracle wires using input-independent SWAP operations.}
\[
O_x^{(R_j,J_j)}
|r_j\rangle_{R_j}|i_j\rangle_{J_j}
=
x_{i_j}^{r_j}
|r_j\rangle_{R_j}|i_j\rangle_{J_j}.
\]
It acts as the identity on all other registers.
On the address branch $\mathbf i$, the oracle acts on
$R_j$ as $Z^{y_j(\mathbf i)}$.
Since $HZ^{y_j(\mathbf i)}H=X^{y_j(\mathbf i)}$, we have
\be
L_x \ket{\mathbf i}_J \ket{r}_R \ket{d}_D
=
\ket{\mathbf i}_J \ket{r\oplus y(\mathbf i)}_R \ket{d}_D.
\ee
Thus, $L_x$ uses $t$ oracle queries and adds the sampled
bit string to the mask register, without requiring the
mask to be initialized to zero.

Next, for $s\in\{0,1\}^t$, let $e_s\in\{0,1\}^{2^t}$
be the string whose only nonzero entry is at position $s$.
Define the input-independent unitary
\be
B \ket{r}_R \ket{d}_D = \ket{r}_R \ket{d\oplus e_r}_D .
\ee
This operation flips the table bit indexed by the current
mask value. For a fixed address branch, writing
$y=y(\mathbf i)$, the sequence $L_xBL_x$ acts as
\be
L_xBL_x\ket{\mathbf i}_J \ket{r}_R \ket{d}_D
=
\ket{\mathbf i}_J \ket{r}_R \ket{d\oplus e_{r\oplus y}}_D.
\ee
It therefore restores the mask register and flips
exactly one table bit, at position $r\oplus y$.

To extract $g(y)$ from this change, define
\[
\Phi_g(r,d)
=
\bigoplus_{s\in\{0,1\}^t}
d_s\,g(r\oplus s) \in\{0,1\},
\]
and the input-independent unitary
\be
A_g
\ket{b}_{\text{clean}} \ket{\mathbf i}_J \ket{r}_R \ket{d}_D 
=
\ket{b\oplus\Phi_g(r,d)}_{\text{clean}} \ket{\mathbf i}_J \ket{r}_R \ket{d}_D .
\ee
For fixed $r$, the function $\Phi_g(r,d)$ is linear in
the table bits $d_s$ over $\mathbb F_2$.
Consequently, flipping the table bit at position
$r\oplus y$ changes $\Phi_g$ by exactly
$g(r\oplus(r\oplus y))=g(y)$.
This motivates the choice
\be
U_x=A_gL_xBL_xA_g.
\ee
To verify its action explicitly, fix $\mathbf i,r,d$,
write $y=y(\mathbf i)$, and let
$d'=d\oplus e_{r\oplus y}$.
The two applications of $A_g$ change the output bit by
\begin{align*}
\Phi_g(r,d)\oplus\Phi_g(r,d')
=
\bigoplus_{s\in\{0,1\}^t}
(d_s\oplus d'_s)\,g(r\oplus s)
=g(r\oplus(r\oplus y))
=g(y).
\end{align*}
Hence,
\be
U_x|b,\mathbf i,r,d\rangle
=
|b\oplus g(y(\mathbf i)),\mathbf i,r,
d\oplus e_{r\oplus y(\mathbf i)}\rangle.
\ee
In particular, the change in the output bit is independent
of both $r$ and $d$. Although the table is not restored
to its original basis value, this does not affect the
output.

\paragraph{Final measurement.}
The initial output bit is $b=0$.
The preceding identity therefore implies that the
reduced state of the clean qubit after applying $U_x$ is
\[
\operatorname{Tr}_{JRD}(U_x\rho_0U_x^\dagger)
=
\frac{1}{N^t}
\sum_{\mathbf i\in[N]^t}
|g(y(\mathbf i))\rangle
\langle g(y(\mathbf i))|.
\]
Thus, measuring the clean qubit gives
\[
\Pr[\text{output}=1]
=
\frac{1}{N^t}
\sum_{\mathbf i\in[N]^t}g(y(\mathbf i))
=
\Pr[g(Y_1,\ldots,Y_t)=1],
\]
as required.

Only the two applications of $L_x$ invoke the input
oracle, so the total number of queries is $2t$.
All other operations are input-independent unitaries,
and no intermediate measurements, resets, or additional
clean qubits are used.
\end{proof}

\begin{remark}
The construction is intended purely as a query complexity simulation. It uses one clean output qubit together with
$t\log_2N+t+2^t$
maximally mixed workspace qubits, including a \(2^t\)-qubit table register, and may involve exponentially large input-independent reversible operations. No intermediate measurements, resets, or additional clean qubits are required. Thus the construction is valid in our DQC1 query model, which places no restriction on mixed workspace or on the complexity of input-independent unitaries, but it does not provide a polynomial-space or polynomial-gate implementation.
\end{remark}

\begin{proof}[Proof of Proposition \ref{prop:Quadratic query comparison}]
By Lemma~\ref{lem:classical-sampling-symmetric}, there exists a classical algorithm that computes $f$ with error probability at most $\varepsilon$ using
$t=\lceil {4\Lambda(F)}/{\varepsilon} \rceil$
independent uniform samples of the input coordinates. Its output depends only on the sampled bits $Y_1,\ldots,Y_t$ and can therefore be written as $(-1)^{g(Y_1,\ldots,Y_t)}$ for a fixed Boolean function $g:\{0,1\}^t\to\{0,1\}$.
By Lemma~\ref{lem:dqc1-sampling}, there exists a DQC1
algorithm using $2t$ oracle queries whose clean-qubit
measurement has exactly the same distribution as
$g(Y_1,\ldots,Y_t)$.
Interpreting measurement outcome $b\in\{0,1\}$ as
$(-1)^b$, the DQC1 algorithm therefore has the same
output distribution as the classical algorithm.
In particular, its error probability is at most
$\varepsilon$ on every promised input. Hence,
$Q^{\mathrm{DQC1}}_\varepsilon(f)
\le 2t$,
which proves~\eqref{eq:explicit-upper}.
\end{proof}

\section{Average-case classical simulation of DQC1}
\label{sec:dqc1-average-case-simulation}

A longstanding folklore conjecture in quantum query complexity asserted that the acceptance probability of any $T$-query
quantum algorithm can be approximated on most uniformly random inputs by a classical algorithm making only $\operatorname{poly}(T)$
queries. Aaronson and Ambainis~\cite{AaronsonAmbainis2014}
showed that this conjecture follows from their influential-variable conjecture for bounded low-degree polynomials.
The latter conjecture was recently proved by Turturean~\cite{turturean2026proof}, thereby resolving the folklore simulation conjecture.

Here we establish stronger quantitative simulation guarantees
for the DQC1 query model, applicable to arbitrary product distributions and all exchangeable distributions.
Remarkably, our proofs are elementary, relying only on the coordinate-sensitivity bound from Theorem~\ref{thm:dqc1-derivative}, together with standard concentration and coupling arguments.
In particular, they do not require the polynomial degree bound or an influential-variable argument.
This illustrates how the additional structural constraints of DQC1 lead directly to strong average-case classical simulation guarantees.

Let $\mathcal{A}$ be a $T$-query DQC1 algorithm, and let $p_\mathcal{A}:\{-1,1\}^N\to[0,1]$ denote its acceptance probability. Choosing the
clean-qubit outcome $0$ as acceptance, we have $p_\mathcal{A}(x)=(1+p(x))/2$, where $p(x)$ is the output bias in \eqref{eq:dqc1-output-bias}.
It follows from Theorem~\ref{thm:dqc1-derivative} that, for any inputs $x,y$ differing in one coordinate,
\begin{equation}
    |p_\mathcal{A}(x)-p_\mathcal{A}(y)|\le \frac{T}{N}.
    \label{eq:avg-simulation-coordinate-bound}
\end{equation}
Consequently, by telescoping along a shortest Hamming path,
\begin{equation}
    |p_\mathcal{A}(x)-p_\mathcal{A}(y)|\le \frac{T}{N}d_H(x,y),
    \qquad x,y\in\{-1,1\}^N.
    \label{eq:avg-simulation-hamming-Lipschitz}
\end{equation}
In what follows, the classical simulator is required to approximate $p_\mathcal{A}(x)$, rather than to reproduce the entire quantum computation.  Only the number of oracle queries is bounded; its preprocessing time and computational complexity are unrestricted.

Our first result is an immediate consequence of bounded differences. Importantly, it uses only~\eqref{eq:avg-simulation-coordinate-bound}, and does not require the algebraic degree bound on $p_A$.

\begin{theorem}[Classical simulation under product distributions]
\label{thm:dqc1-product-simulation}
Let $\mathcal A$ be a $T$-query DQC1 algorithm with $T\ge 1$,
and let $\mu=\mu_1\otimes\cdots\otimes\mu_N$ be any product
distribution on $\{-1,1\}^N$. Then, for every $s>0$,
\begin{equation}
\Pr_{X\sim\mu}\Big[
\big|p_{\mathcal A}(X)
-\mathbb E_\mu[p_{\mathcal A}(X)]\big|\ge s
\Big]
\le
2\exp\left(-\frac{2Ns^2}{T^2}\right).
\label{eq:dqc1-product-concentration}
\end{equation}
Moreover, for every $\varepsilon,\delta\in(0,1)$, there
exists a deterministic, nonadaptive classical query algorithm
$\mathcal{C}:\{-1,1\}^N\to[0,1]$, depending on
$\mathcal A,\mu,\varepsilon,\delta$, such that
\begin{equation}
\Pr_{X\sim\mu}\left[
|\mathcal{C}(X)-p_{\mathcal A}(X)|>\varepsilon
\right]\le\delta.
\end{equation}
The number of classical queries is at most
\begin{equation}
\min\left\{
N,\left\lceil
\frac{T^2}{2\varepsilon^2}
\log\frac{2}{\delta}
\right\rceil
\right\}.
\label{eq:dqc1-product-query-bound}
\end{equation}
In particular, if
$N\ge \left\lceil
T^2(2\varepsilon^2)^{-1}\log(2/\delta)
\right\rceil$,
the simulator makes no queries.
\end{theorem}

\begin{proof}
By~\eqref{eq:avg-simulation-coordinate-bound}, changing the $i$th coordinate changes $p_\mathcal{A}$ by at most $T/N$. Since the coordinates of $X\sim\mu$ are independent, McDiarmid's bounded differences inequality~\cite{McDiarmid1989} gives
\[
\Pr_{X\sim\mu}\Big[
 |p_\mathcal{A}(X)-\mathbb E_\mu [p_\mathcal{A}(X)]|\ge s \Big]
\le 2\exp\!\left(-\frac{2 s^2}{N(T/N)^2}\right),
\]
which proves~\eqref{eq:dqc1-product-concentration}.
For classical simulation, set $K=\lceil T^2(2\varepsilon^2)^{-1}\log(2/\delta)\rceil$.
If $N\ge K$, the simulator makes no queries and simply outputs $\mathcal{C}(x)=\mathbb E_\mu[p_{\mathcal A}(X)].$ 
If $N<K$, the simulator queries all $N$ coordinates and outputs $p_{\mathcal A}(x)$ exactly. Both simulators are deterministic and nonadaptive, and the claimed query bound follows.
\end{proof}

Thus every $T=o(\sqrt N)$ DQC1 query algorithm has an acceptance probability converging in probability to a constant under any sequence of product input distributions, for any fixed error threshold. In particular, a $\operatorname{poly}(\log N)$-query DQC1 algorithm is asymptotically indistinguishable from a constant-output algorithm on a typical product-distributed input. This is an average-case statement and does not rule out speedups on structured promise domains. The corresponding second-moment bound is also useful. Let \(a_i\) denote the probability that the \(i\)-th input coordinate takes the value \(-1\) under the distribution \(\mu\). The tensorization of variance and
\eqref{eq:avg-simulation-coordinate-bound} give
\begin{equation}\label{eq:sec8-product-variance}
\operatorname{Var}_{\mu}(p_\mathcal{A}(X))
\le \frac{T^2}{N^2}\sum_{i=1}^N a_i(1-a_i)
\le \frac{T^2}{4N}.
\end{equation}
%In particular, the uniform-distribution variance estimate obtained from Proposition~\ref{prop:fourier-concentration} extends to all product distributions.

\begin{remark}
The simulations in Theorem \ref{thm:dqc1-product-simulation} rely on the uniform coordinate-sensitivity bound, rather than on the polynomial degree bound alone. The tracial structure of DQC1 limits the change in acceptance probability under any single-coordinate flip to $T/N$. When $T=o(\sqrt{N})$, this forces concentration under product distributions and within Hamming slices, providing the basis for our classical simulations.
In contrast, standard quantum query algorithms can concentrate their query weight on individual input coordinates, and no analogous $O(T/N)$ sensitivity bound holds in general. Consequently, low polynomial degree alone does not guarantee the concentration needed for our argument. 
\end{remark}

The product-distribution assumption in Theorem \ref{thm:dqc1-product-simulation} is not necessary for concentration. We below consider a more general case. For $0\le k\le N$, define the Hamming slice
\[
\Omega_k=\{x\in\{-1,1\}^N:\operatorname{wt}(x)=k\},
\quad |\Omega_k|=\binom{N}{k},
\]
and let $\nu_k$ denote the uniform distribution on $\Omega_k$. Write
\[
m_k= \mathop{\mathbb{E}}\limits_{X\sim\nu_k} [p_\mathcal{A}(X)] = \frac{1}{\binom{N}{k}}
\sum_{x\in\Omega_k} p_\mathcal{A}(x)
\]
as the average acceptance probability of $\mathcal{A}$ over the corresponding Hamming slice $\Omega_k$. Concentration inequalities on Hamming slices and, more generally, multislices have been extensively studied; see, e.g., \cite{SambaleSinulis2022}.
In our setting, the coordinate-sensitivity bound for DQC1 query algorithms leads to the following concentration inequality via a simple martingale argument.

\begin{lemma}%[Concentration on a Hamming slice]
\label{lem:dqc1-slice-concentration}
Let $\mathcal{A}$ be a $T$-query DQC1 algorithm with $T\ge1$, and put $r=\min\{k,N-k\}$. If $1\le r\le N/2$, then for every $s>0$,
\begin{equation}
\Pr_{X\sim\nu_k} \left[|p_\mathcal{A}(X)-m_k|\ge s\right]
\le 2\exp\!\left(-\frac{N^2 s^2}{8rT^2}\right)
\le 2\exp\!\left(-\frac{N s^2}{4T^2}\right).
\label{eq:dqc1-slice-concentration}
\end{equation}
Moreover,
\begin{equation}\label{eq:sec8-slice-variance}
\operatorname{Var}_{\nu_k}(p_\mathcal{A} (X))
\le \frac{rT^2}{N^2}\le \frac{T^2}{2N}.
\end{equation}
For $k\in\{0,N\}$, the slice contains a single input and $p_\mathcal{A}(X)=m_k$ identically.
\end{lemma}

\begin{proof}
By symmetry between the two bit values, a uniform point of $\Omega_k$ can be generated by choosing $r=\min\{k,N-k\}$ positions uniformly without replacement, assigning them the value $-1$ if $k\le N/2$ and $+1$ otherwise, and assigning the opposite value to all remaining positions.
Expose the chosen positions sequentially as $Z_1,\ldots,Z_r$, and let
\[
M_j=\mathbb E\left[p_\mathcal{A}(X)\mid Z_1,\ldots,Z_j\right],
\qquad 0\le j\le r,
\]
be the corresponding Doob martingale. In particular, $M_0=m_k$ and $M_r=p_\mathcal{A}(X)$.

We aim to show that $|M_j-M_{j-1}|\le 2T/N$. To this end, it suffices to show that, conditioned on $Z_1,\ldots,Z_{j-1}$, the conditional expectations corresponding to any two possible values of $Z_j$ differ by at most $2T/N$, since $M_{j-1}$ is their weighted average.
Fix $Z_1,\ldots,Z_{j-1}$ and consider two distinct remaining indices $a,b$ as possible values of $Z_j$. The remaining uniformly sampled positions can be coupled by transposing $a$ and $b$, so that the two completed strings differ in at most two coordinates. By~\eqref{eq:avg-simulation-hamming-Lipschitz}, their acceptance probabilities differ by at most $2T/N$. Taking expectations under this coupling gives the desired bound.
Azuma--Hoeffding's inequality now gives
\[
\Pr[|M_r-M_0|\ge s]
\le2\exp\!\left(-\frac{s^2}{2r(2T/N)^2}\right)
=2\exp\!\left(-\frac{N^2s^2}{8rT^2}\right)
\]
proving \eqref{eq:dqc1-slice-concentration}. Furthermore, conditional on the previously exposed positions, the variance of $M_j$ is at most $(2T/N)^2/4=T^2/N^2$. Orthogonality of the martingale differences gives
\[
\operatorname{Var}(M_r)
=\sum_{j=1}^r\mathbb E\left[(M_j-M_{j-1})^2\right]
\le\frac{rT^2}{N^2},
\]
which proves~\eqref{eq:sec8-slice-variance}.
\end{proof}

In particular, if the slice index $k$ is known and $N$ is sufficiently large compared with $T^2$, the constant $m_k$ can be used to approximate the acceptance probability on almost every input without making any queries. The next observation describes how these constants vary as $k$ changes.

\begin{lemma}%[Lipschitz continuity of slice averages]
\label{lem:dqc1-slice-means}
For $0\le k<N$, we have $|m_{k+1}-m_k| \leq T/N$.
Consequently, the piecewise-linear interpolation $M:[0,1]\to[0,1]$ specified by $M(k/N)=m_k$ has Lipschitz constant at most $T$.
\end{lemma}

\begin{proof}
Sample $X\sim\nu_k$ and choose uniformly one of its $N-k$ coordinates equal to $+1$. Flip that coordinate to $-1$, obtaining $Y\sim\nu_{k+1}$. The coupled inputs differ in exactly one coordinate, so~\eqref{eq:avg-simulation-coordinate-bound} gives
$|m_{k+1}-m_k|
=|\mathbb E[p_\mathcal{A}(Y)-p_\mathcal{A}(X)]|
\le T/N.$
The statement about $M$ follows by linear interpolation.
\end{proof}

A distribution $\mu$ on $\{-1,1\}^N$ is \emph{exchangeable} if it is invariant under every permutation of the input coordinates~\cite{DiaconisFreedman1980}. 
Equivalently, it is a mixture of uniform Hamming slices,
\begin{equation}
\mu=\sum_{k=0}^N w_k\nu_k,
\qquad w_k\ge0,\qquad \sum_{k=0}^Nw_k=1.
\label{eq:dqc1-exchangeable-mixture}
\end{equation}
The preceding two lemmas imply an approximate symmetry
property of DQC1 acceptance probabilities.
Although the acceptance probability of a DQC1 algorithm
need not be symmetric, it can be approximated under
any exchangeable input distribution by a Lipschitz
function of the normalized Hamming weight.
Moreover, the same approximating function works for
all exchangeable distributions, as shown below.

\begin{corollary}
\label{cor:approximate-symmetry}
Let $\mathcal{A}$ be a $T$-query DQC1 algorithm with $T\ge 1$, and let $p_\mathcal{A}:\{-1,1\}^N\to[0,1]$ denote its acceptance probability. Then there exists a $T$-Lipschitz function $M:[0,1]\to[0,1]$ such that, for every exchangeable distribution $\mu$ on $\{-1,1\}^N$ and every $s>0$,
\begin{equation}
\Pr_{X\sim\mu}\left[\,
\left|p_\mathcal{A}(X)-
M\left(\frac{\operatorname{wt}(X)}{N}\right)\right|
\ge s
\right]
\le
2\exp\left(-\frac{Ns^2}{4T^2}\right).
\label{eq:approximate-symmetry-concentration}
\end{equation}
Moreover,
\begin{equation}
\mathop{\mathbb E}\limits_{X\sim\mu}
\left[
\left(
p_A(X)-
M\left(\frac{\operatorname{wt}(X)}{N}\right)
\right)^2
\right]
\le \frac{T^2}{2N}.
\label{eq:approximate-symmetry-variance}
\end{equation}
The same function $M$ works for all exchangeable distributions
$\mu$.
\end{corollary}

\begin{proof}
We use the notations from Lemmas~\ref{lem:dqc1-slice-concentration} and~\ref{lem:dqc1-slice-means}.
Fix $k\in\{0,\ldots,N\}$. For every $X\in\Omega_k$,
we have $M(\operatorname{wt}(X)/N)=m_k$.
Thus, Lemma~\ref{lem:dqc1-slice-concentration} gives \eqref{eq:approximate-symmetry-concentration} for $X\sim \nu_k$.
Now let $\mu$ be any exchangeable distribution of the form \eqref{eq:dqc1-exchangeable-mixture}.
Averaging the preceding concentration inequality over $k$
proves~\eqref{eq:approximate-symmetry-concentration}.
For the second-moment bound, observe that
\beas
\mathop{\mathbb E}\limits_{X\sim\mu}
\left[
\left(
p_\mathcal{A}(X)-
M\left(\frac{\operatorname{wt}(X)}{N}\right)
\right)^2
\right] 
&=&
\sum_{k=0}^N w_k
\mathop{\mathbb E}\limits_{X\sim\nu_k}
\left[(p_\mathcal{A}(X)-m_k)^2\right] \\
&=&
\sum_{k=0}^N w_k
\operatorname{Var}_{\nu_k}(p_\mathcal{A}(X)) \\
&\le&
\sum_{k=0}^N w_k\frac{T^2}{2N}
=\frac{T^2}{2N},
\eeas
where the inequality follows from the variance bound
in Lemma~\ref{lem:dqc1-slice-concentration}.
\end{proof}

Corollary~\ref{cor:approximate-symmetry} shows that,
when $T=o(\sqrt{N})$, the acceptance probability of a
DQC1 algorithm is asymptotically close to a symmetric
function under every exchangeable input distribution,
both in probability and in $L^2$.
Importantly, this is a distributional approximation
rather than a pointwise one, and does not require the
original DQC1 algorithm to have any symmetry.
Equivalently, let $\overline p_\mathcal{A}$ denote the permutation
symmetrization of $p_\mathcal{A}$, defined by
\[
\overline p_\mathcal{A}(x)
=
\frac{1}{N!}\sum_{\pi\in S_N}
p_\mathcal{A}(x_{\pi(1)},\ldots,x_{\pi(N)}).
\]
Since
$\overline p_\mathcal{A}(x)=M(\operatorname{wt}(x)/N)$,
Corollary~\ref{cor:approximate-symmetry} implies that
$\|p_\mathcal{A}-\overline p_\mathcal{A}\|_{L^2(\mu)}
\le {T}/{\sqrt{2N}}$
for every exchangeable distribution $\mu$.
Thus, low-query DQC1 acceptance probabilities are
approximately invariant under coordinate permutations
in an average-case sense.

The approximate symmetry property established above
also leads to an efficient average-case classical simulation.
Although the acceptance probability need not concentrate
around a single constant under an exchangeable distribution,
it is well approximated by a Lipschitz function of the
normalized Hamming weight. By estimating this weight
from a small sample of input coordinates, we obtain a
deterministic, nonadaptive classical simulation that works
simultaneously for all exchangeable distributions.

\begin{theorem}[Classical simulation under exchangeable distributions]
\label{thm:dqc1-exchangeable-simulation}
Let $\mathcal A$ be a $T$-query DQC1 algorithm.
For every $\varepsilon,\delta\in(0,1)$, there exists a
deterministic, nonadaptive classical query algorithm
$\mathcal{C}:\{-1,1\}^N\to[0,1]$, depending on
$\mathcal A,\varepsilon,\delta$ but not on the input
distribution, such that
\begin{equation}
\Pr_{X\sim\mu}
\big[ |\mathcal{C}(X)-p_{\mathcal A}(X)|>\varepsilon \big]
\le\delta
\label{eq:dqc1-uniform-slice-simulation}
\end{equation}
for every exchangeable distribution $\mu$.
Moreover, $\mathcal{C}$ makes at most
\begin{equation}
\min\left\{
N,\left\lceil
\frac{16T^2}{\varepsilon^2}
\log\frac{4}{\delta}
\right\rceil
\right\}
\label{eq:dqc1-exchangeable-query-bound}
\end{equation}
classical queries.
\end{theorem}

\begin{proof}
The case $T=0$ is immediate. Suppose $T\ge1$, and set
$q=\lceil 16 T^2 \varepsilon^{-2} \log(4/\delta) \rceil.$
If $q\ge N$, the simulator queries all $N$ coordinates
and outputs $p_{\mathcal A}(x)$ exactly.
Hence, assume $q<N$.
Let $M:[0,1]\to[0,1]$ be the $T$-Lipschitz function
given by Corollary~\ref{cor:approximate-symmetry}.
The simulator queries the first $q$ coordinates and outputs
\[
\mathcal{C} (x)=M\left(
\frac1q\sum_{i=1}^q\frac{1-x_i}{2}
\right).
\]
Clearly, $\mathcal{C}$ is deterministic and nonadaptive.
Fix $k\in\{0,\ldots,N\}$ and let $X\sim\nu_k$.
Write
\[
\theta=\frac{k}{N},
\qquad
\widehat\theta=
\frac1q\sum_{i=1}^q\frac{1-X_i}{2}.
\]
By Corollary~\ref{cor:approximate-symmetry},
\[
\Pr\left[
|p_{\mathcal A}(X)-M(\theta)|
>\frac{\varepsilon}{2}
\right]
\le
2\exp\left(
-\frac{N\varepsilon^2}{16T^2}
\right)
\le\frac{\delta}{2}.
\]
Moreover, Hoeffding's inequality for sampling
without replacement gives
\[
\Pr\left[
|\widehat\theta-\theta|
>\frac{\varepsilon}{2T}
\right]
\le
2\exp\left(
-\frac{q\varepsilon^2}{2T^2}
\right)
\le\frac{\delta}{2}.
\]
Outside these two exceptional events, the
$T$-Lipschitz continuity of $M$ implies
\begin{align*}
|\mathcal{C}(X)-p_{\mathcal A}(X)|
\le
|M(\widehat\theta)-M(\theta)|
+|M(\theta)-p_{\mathcal A}(X)|
\le
T|\widehat\theta-\theta|
+\frac{\varepsilon}{2}
\le\varepsilon.
\end{align*}
Thus, by the union bound, $\Pr_{X\sim\nu_k}
\left[
|\mathcal{C}(X)-p_{\mathcal A}(X)|>\varepsilon
\right]
\le\delta.$
Since the same simulator works for every $k$,
the conclusion extends to all exchangeable
distributions by~\eqref{eq:dqc1-exchangeable-mixture}.
\end{proof}

% The results above concern query complexity alone. In particular, the simulator may use precomputed rational approximations to the relevant expectations, and no bound is claimed on its running time or description size. Under product distributions the simulator may also depend on the distribution; in contrast, Theorem~\ref{thm:dqc1-exchangeable-simulation} provides one simulator that works for all exchangeable distributions simultaneously. The conclusions do not extend automatically to arbitrary correlated promise distributions.

% For uniformly random inputs, Proposition~\ref{prop:fourier-concentration}  implies $\operatorname{Var}(p_A(X))\le T^2/4N$. Thus, Chebyshev's inequality alone gives a polynomial-query average-case simulation under the uniform distribution. The bounded-differences argument above gives an exponential concentration tail, the logarithmic dependence on $1/\delta$, and extensions beyond independent unbiased inputs. More generally, these results show that the coordinate-sensitivity constraint is useful not only for query lower bounds but also for identifying input distributions on which DQC1 output statistics admit simple classical approximations.

\begin{remark}
Theorems~\ref{thm:dqc1-product-simulation} and
\ref{thm:dqc1-exchangeable-simulation} provide a particularly
simple and quantitative form of average-case classical
simulation for DQC1 algorithms. In light of the recent
resolution of the Aaronson-Ambainis conjecture
\cite{turturean2026proof}, polynomial-query average-case classical
simulation is now known for general BQP query algorithms
under the uniform input distribution.
Our results show that, in the DQC1 model, explicit quadratic
query overhead and deterministic, nonadaptive simulation
follow directly from the coordinate-sensitivity constraint,
without requiring the polynomial degree bound or an
influential-variable argument.
Moreover, these guarantees extend to arbitrary product and
exchangeable input distributions, with a distribution-independent
simulator in the latter case.
The approximate symmetry property in
Corollary~\ref{cor:approximate-symmetry} further illustrates
the strong structural restrictions on DQC1 acceptance
probabilities.

% These results also constrain possible quantum--classical
% query separations. For any promise problem admitting a
% $T$-query bounded-error DQC1 algorithm, every product or
% exchangeable input distribution supported on the promise
% admits a deterministic, nonadaptive classical algorithm
% using $O(T^2)$ queries and achieving constant average error.
% Hence, such distributions cannot witness superpolynomial
% DQC1--classical query separations.
% In particular, since exchangeable distributions may exhibit
% strong correlations among input coordinates, correlations
% alone are insufficient to explain large DQC1 query
% advantages. This observation is consistent with known
% exponential query separations for structured problems such
% as cyclic Forrelation~\cite{zhang-shao,ji2026dqc1},
% and raises the question of what additional structure in
% input distributions enables such separations.
% Understanding this structure, potentially through stronger
% constraints on DQC1 output polynomials beyond coordinate
% sensitivity, remains an interesting direction for future work.
\end{remark}

\section{Fourier consequences of coordinate sensitivity}
\label{sec:Fourier analysis}

The Fourier structure of DQC1 query algorithms has been studied in considerably greater detail in recent work. Girish~\cite{Girish2026Fourier} obtained Fourier growth bounds for the acceptance probabilities of $\mathrm{DQC}k$. Later, Escudero Guti\'errez, Saucedo, and Palazuelos~\cite{gutierrez2026optimal} established
improved Fourier-growth bounds at the highest and
second-highest levels of the acceptance probability polynomial.

Our purpose here is different. Rather than seeking to improve these bounds, we ask how much Fourier structure is already forced by the much simpler polynomial relaxation introduced in this work, namely low algebraic degree together with uniformly bounded coordinate sensitivity. We show that these two constraints alone imply nontrivial Fourier concentration and fixed level Fourier growth bounds, uniformly under restrictions. 
Although these bounds are generally less refined than those known for actual DQC1 output polynomials, they apply to the broader class of sensitivity-constrained polynomials and isolate the Fourier suppression captured by coordinate sensitivity alone.
As an illustration, this elementary viewpoint is sufficient to prove a lower bound of $N^{1/4}$ for the sensitivity approximate degree of 2-Forrelation problem, which in turn yields the corresponding \(N^{1/4}\) lower bound on DQC1 query complexity for 2-Forrelation~\cite{Girish2026Fourier}.

\begin{proposition}
\label{prop:fourier-concentration}
Let $p(x)=\sum_{S\subseteq[N]}\widehat p(S)\chi_S(x)$
be a real multilinear polynomial satisfying
\be
\label{condition}
\deg p\le d,
\qquad
\|p\|_\infty\le 1,
\qquad
\max_{i\in[N]}\|\partial_i p\|_\infty
\le \frac{d}{2N}.
\ee
Then
\begin{equation}
\sum_{S\subseteq[N]} |S|\,\widehat p(S)^2
\le
\frac{d^2}{4N}.
\label{eq:fourier-weighted-l2}
\end{equation}
% Consequently, for every integer $k\ge1$,
% \begin{equation}
% \sum_{\substack{S\subseteq[N]\\ |S|\ge k}}
% \widehat p(S)^2
% \le
% \frac{d^2}{4Nk}.
% \label{eq:fourier-tail-l2}
% \end{equation}
\end{proposition}

\begin{proof}
By Parseval's identity and the standard Fourier formula for Boolean derivatives \cite{ODonnell2014}, we have
$\sum_{S\subseteq[N]} |S|\,\widehat p(S)^2
=
\sum_{i=1}^N
\|\partial_i p\|_2^2.$
Using the coordinate-sensitivity bound, we obtain
\[
\sum_{i=1}^N\|\partial_i p\|_2^2
\le
\sum_{i=1}^N\|\partial_i p\|_\infty^2
\le
N\left(\frac{d}{2N}\right)^2
=
\frac{d^2}{4N}.
\]
This proves \eqref{eq:fourier-weighted-l2}. 
% The second bound follows from
% $k
% \sum_{|S|\ge k}\widehat p(S)^2
% \le
% \sum_S |S|\,\widehat p(S)^2.$
\end{proof}

In particular, when $k=1$, the variance
$\operatorname{Var}(p)
=
\sum_{S\ne\emptyset}\widehat p(S)^2
\le
{d^2}/{4N}.$
Thus, when $d=o(\sqrt N)$, such a polynomial is asymptotically close to a
constant in $L_2$ under the uniform distribution.

The preceding proposition bounds the Fourier
spectrum in the $L^2$ norm. We next investigate
the corresponding $L^1$ bounds, which are
particularly useful in Fourier-analytic
query-complexity lower bounds.
For an integer $\ell\geq1$, define the
level-$\ell$ Fourier growth of $p$ by
\[
L_{1,\ell}(p)
:=
\sum_{ S\subseteq[N], |S|=\ell} 
|\widehat p(S)|.
\]
The following counting identity relates the
Fourier growth of $p$ to that of its coordinate
derivatives:
\begin{equation}
\label{eq:fourier-growth-derivative-identity}
\ell L_{1,\ell}(p)
=
\sum_{i=1}^N
L_{1,\ell-1}(\partial_i p),
\qquad \ell\geq1,
\end{equation}
where
$L_{1,0}(q)=|\widehat q(\varnothing)|$.
Indeed,
$\partial_i p(x)
=
\sum_{S\ni i}
\widehat p(S)\chi_{S\setminus\{i\}}(x),$
and each Fourier coefficient at level $\ell$
is counted exactly $\ell$ times on the
right-hand side of
\eqref{eq:fourier-growth-derivative-identity}.

In particular, the coordinate sensitivity
constraint immediately gives
\begin{equation}
\label{eq:fourier-level-one}
L_{1,1}(p)
=
\sum_{i=1}^N
\left|\mathbb E_x[\partial_i p(x)]\right|
\leq
\sum_{i=1}^N\|\partial_i p\|_\infty
\leq d.
\end{equation}
To obtain a stronger bound at level two,
we use an elementary first-level Fourier
inequality for bounded low-degree polynomials.

\begin{lemma}%[First-level Fourier bound]
\label{lem:first-level-fourier-bound}
Let $q:\{-1,1\}^m\to\mathbb R$ be a
multilinear polynomial of degree at most
$r\geq1$. Then
$L_{1,1}(q)\leq r\|q\|_\infty.$
\end{lemma}

\begin{proof}
Choose signs $s_i\in\{-1,1\}$ such that
$s_i\widehat q(\{i\})
=
|\widehat q(\{i\})|.$
For $t\in[-1,1]$, let $T_t$ denote the
Boolean noise operator, defined by
$
T_t q(x)
=
\sum_{S\subseteq[m]}
t^{|S|}\widehat q(S)\chi_S(x).
$
Consider the univariate polynomial $g(t)=(T_tq)(s).$
It has degree at most $r$. Moreover,
$T_t$ is an $L^\infty$ contraction for
every $t\in[-1,1]$, so $\|g\|_{L^\infty([-1,1])}
\leq\|q\|_\infty.$
By Bernstein's inequality at $t=0$, $|g'(0)|
\leq r\|g\|_{L^\infty([-1,1])}
\leq r\|q\|_\infty.$
In addition,
$g'(0)
=
\sum_{i=1}^m
s_i\widehat q(\{i\})
=
L_{1,1}(q),$
which proves the claim.
\end{proof}

\begin{proposition}%[Level-two Fourier growth]
\label{prop:level-two-fourier-growth}
Let $p:\{-1,1\}^N\to\mathbb R$ be a real multilinear polynomial satisfying \eqref{condition}.
Then $L_{1,2}(p)
\le
{d(d-1)}/{4}.$
More generally, the same bound holds for every restriction $\rho$.
\end{proposition}

\begin{proof}
From the Fourier expansion of the Boolean derivative, $\partial_i p=\sum_{S\ni i}\widehat p(S)\chi_{S\setminus\{i\}},$
and hence, by counting each level-two coefficient twice,
\begin{equation}
2L_{1,2}(p)
=
\sum_{i=1}^N L_{1,1}(\partial_i p).
\label{eq:derivative-fourier-counting}
\end{equation}
By Lemma \ref{lem:first-level-fourier-bound},
$L_{1,1}(q)
\le
r\|q\|_\infty$
for every multilinear polynomial $q$ of degree at most $r$. Applying this
to $q=\partial_i p$, whose degree is at most $d-1$, gives
\[
L_{1,1}(\partial_i p)
\le
(d-1)\|\partial_i p\|_\infty
\le
\frac{d(d-1)}{2N}.
\]
Substituting into \eqref{eq:derivative-fourier-counting},
\[
L_{1,2}(p)
\le
\frac{N}{2} \frac{d(d-1)}{2N}
=
\frac{d(d-1)}4.
\]
Now let $\rho$ be a restriction leaving $m\le N$ variables free. Then
$\deg(p_\rho)\le d,$ and $\max_i\|\partial_i(p_\rho)\|_\infty
\le
{d}/{2N}.$
Repeating the same argument proves the second claim.
\end{proof}

The same derivative-counting identity, combined with known Fourier-growth inequalities for bounded low-degree polynomials, yields bounds at every fixed Fourier level.

\begin{proposition}
\label{prop:fixed-level-fourier-growth}
Let $p:\{-1,1\}^N\to\mathbb R$ be a real multilinear polynomial satisfying \eqref{condition}.
Then for every fixed integer $\ell\geq2$,
we have $L_{1,\ell}(p)
=
O_\ell (
d^\ell N^{(\ell-2)/2}
).$
The same bound holds for every restriction
of $p$, with the original ambient dimension
$N$ on the right-hand side.
\end{proposition}

\begin{proof}
We use the Fourier-growth inequality for
bounded low-degree polynomials due to
Iyer et al.~\cite{IyerRaoReisRothvossYehudayoff2021}:
for every fixed integer $j\geq1$,
a polynomial $q$ on $m$ Boolean variables
of degree at most $r$ satisfies
$L_{1,j}(q)
\leq
C_j r^j m^{(j-1)/2}\|q\|_\infty,$
where $C_j$ depends only on $j$.
Apply this inequality with $j=\ell-1$
to $q_i=\partial_i p$.
Since $\deg q_i\leq d-1$ and $\|q_i\|_\infty\leq {d}/{2N},$
we obtain
\[
L_{1,\ell-1}(\partial_i p)
\leq
C_{\ell-1}
(d-1)^{\ell-1}
N^{(\ell-2)/2}
\frac{d}{2N}.
\]
Summing over $i$ and applying
\eqref{eq:fourier-growth-derivative-identity},
\beas
L_{1,\ell}(p)
=
\frac1\ell\sum_{i=1}^N
L_{1,\ell-1}(\partial_i p) 
\leq
\frac{C_{\ell-1}}{2\ell}
d(d-1)^{\ell-1}
N^{(\ell-2)/2}
=
O_\ell\left(
d^\ell N^{(\ell-2)/2}
\right).
\eeas
Under a restriction, the number of free
variables cannot increase, and each
surviving coordinate derivative retains
the bound $d/2N$. So the same argument
applies uniformly to every restriction.
\end{proof}

\begin{remark}
For comparison, the general Fourier-growth bound for an arbitrary bounded
degree-$d$ polynomial \cite[Theorem~3]{IyerRaoReisRothvossYehudayoff2021} is 
$L_{1,\ell}(p)
=
O_\ell (
d^\ell N^{(\ell-1)/2}
).$
Thus the coordinate-sensitivity constraint improves the dependence on the
ambient dimension by a factor of $\sqrt N$ at every fixed level
$\ell\ge2$.
For actual DQC1 acceptance probabilities, Girish~\cite[Theorem~1.5]{Girish2026Fourier}
obtained, uniformly under restrictions, the bound
$L_{1,\ell}
=
O\left(C d^3 N^{(\ell-2)/2} \right),$
where $C = c^\ell \sqrt{\ell!} \log^{\ell-2}(N)$
for some constant $c>0$.\footnote{Through email correspondence with Uma Girish, we learned that the bound in \cite[Theorem~1.5]{Girish2026Fourier} can be improved to $O(d^2)$ for $\ell=2$. In this case, Proposition~\ref{prop:level-two-fourier-growth} gives the same dependence through a particularly elementary argument.
Consequently, the lower bound for 2-Forrelation in \cite[Corollary~1.9]{Girish2026Fourier} can be strengthened to $\Omega(N^{1/4})$, matching our bound in Corollary~\ref{cor:two-forrelation-dqc1}.} For fixed $\ell\ge4$, this gives a better polynomial dependence on the number of queries, at the expense of polylogarithmic factors in $N$.
The distinction here is that our bound follows directly from the much simpler polynomial constraints of low degree and bounded coordinate sensitivity, without using the additional matrix structure of DQC1 computations.
\end{remark}

% \subsection{Higher-order derivatives}
% \label{subsec:higher-fourier-derivatives}

% For completeness, the Fourier expansion
% also gives an exact identity for higher-order
% Boolean derivatives. For
% $I\subseteq[N]$, write
% \[
% \partial_I p
% :=
% \prod_{i\in I}\partial_i p.
% \]

% \begin{proposition}[Higher-order derivative identity]
% \label{prop:higher-derivative-identity}
% For every integer $j\geq1$,
% \begin{equation}
% \label{eq:higher-derivative-fourier}
% \sum_{\substack{I\subseteq[N]\\|I|=j}}
% \|\partial_I p\|_2^2
% =
% \sum_{S\subseteq[N]}
% \binom{|S|}{j}\widehat p(S)^2.
% \end{equation}
% Here $\|\cdot\|_2$ is taken with respect
% to the uniform distribution on the
% Boolean cube.
% \end{proposition}

% \begin{proof}
% For every $I\subseteq[N]$, we have
% $\partial_I p
% =
% \sum_{S\supseteq I}
% \widehat p(S)\chi_{S\setminus I}.$
% Parseval's identity therefore gives
% $\|\partial_I p\|_2^2
% =
% \sum_{S\supseteq I}\widehat p(S)^2.$
% Summing over all $I$ of size $j$,
% each Fourier coefficient
% $\widehat p(S)^2$ is counted
% $\binom{|S|}{j}$ times.
% \end{proof}

% \begin{remark}
% The first-order sensitivity constraint
% does not imply an additional factor
% of $1/N$ for each successive derivative.
% Indeed, Boolean differentiation is an
% $L^\infty$ contraction, so for every
% nonempty $I\subseteq[N]$,
% \[
% \|\partial_I p\|_\infty
% \leq
% \max_{i\in I}\|\partial_i p\|_\infty
% \leq\frac{d}{N}.
% \]
% In general, this cannot be improved to
% a bound of order $(d/N)^{|I|}$ without
% additional assumptions.
% \end{remark}

% \subsection{Application to 2-Forrelation}
% \label{subsec:two-forrelation}

We now apply the level-two Fourier-growth bound
to obtain a lower bound for 2-Forrelation.
The key ingredient is the
Fourier growth to Forrelation reduction of
Girish, Raz, and Zhan
\cite[Theorem~3.1 in the full version]{GirishRazZhan2021}.
Unlike a reduction restricted to Boolean-valued
test functions, this result applies to functions
taking values in $[-1,1]$, as required for our
polynomial relaxation. 

We first recall the 2-Forrelation problem. Let $H$ be the 1-qubit Hadamard gate. For $x,y\in\{-1,1\}^N$, define
\[
\Phi_2(x,y)
=
\frac{1}{N} x^{\mathsf T} H^{\otimes n} y.
\]
Let $\varepsilon= (60\log N)^{-1}$.
We consider the partial Boolean function
$\mathrm{Forr}_{2,N}$ on $2N$ variables,
defined by
\[
\mathrm{Forr}_{2,N}(x,y)
=
\begin{cases}
-1,
& \Phi_2(x,y)\geq \varepsilon/2, \\[2mm]
+1,
& \Phi_2(x,y)\leq \varepsilon/4,
\end{cases}
\]
and undefined on inputs outside these two promise cases.
This is the single-copy Forrelation promise used in
\cite{GirishRazZhan2021}.

\begin{theorem}
\label{thm:two-forrelation-sensitivity}
For every fixed $0<\eta<1$, we have
$\dego_\eta(\mathrm{Forr}_{2,N})
=
\Omega_\eta(N^{1/4}).$
\end{theorem}

\begin{proof}
Let $d=\dego_\eta(\mathrm{Forr}_{2,N}),$
and let $p$ be a corresponding approximating polynomial. Since
$\mathrm{Forr}_{2,N}$ has $2N$ input variables,
$\deg p\le d, \|p\|_\infty\le1,$ and $\max_{i\in[2N]}\|\partial_i p\|_\infty
\le {d}/{4N}.$
Let $\mathcal P_d$ denote the restriction closure of the family of
polynomials satisfying these three conditions. By
Proposition~\ref{prop:level-two-fourier-growth},
$L_{1,2}(q)
\le O(d^2)$
for every $q\in\mathcal P_d$. Hence
\begin{equation}
\label{eq:forrelation-family-growth}
L_{1,2}(\mathcal P_d)
:=
\sup_{q\in\mathcal P_d}L_{1,2}(q)
\le O(d^2).
\end{equation}
Let $\mathcal U$ be the uniform distribution on
$\{-1,1\}^{2N}$ and let $\mathcal G$ be the rounded Forrelation
distribution of Girish, Raz, and Zhan
\cite[Definition~2.10]{GirishRazZhan2021}.
Their Fourier-growth reduction
\cite[Theorem~3.1]{GirishRazZhan2021},
specialized to one copy of Forrelation, gives
\[
\left|
\mathbb E_{\mathcal U}[q]
-
\mathbb E_{\mathcal G}[q]
\right|
\le
O\left(
\frac{L_{1,2}(\mathcal P_d)}{\sqrt N}
\right)
+o(N^{-1/2})
\]
for every $q\in\mathcal P_d$. Applying this to $p$ and using
\eqref{eq:forrelation-family-growth},
\begin{equation}
\label{eq:forrelation-advantage-upper}
\left|
\mathbb E_{\mathcal U}[p]
-
\mathbb E_{\mathcal G}[p]
\right|
\le
O\left(\frac{d^2}{\sqrt N}\right)
+o(N^{-1/2}).
\end{equation}
On the other hand,
\cite[Lemma~2.11]{GirishRazZhan2021}
shows that $\mathcal U$ and $\mathcal G$ lie in the $+1$ and $-1$
promise cases, respectively, with probability $1-O(N^{-6})$.
Since $\|p\|_\infty\le1$ and $p$ $\eta$-approximates
$\mathrm{Forr}_{2,N}$ on its promise domain,
$\mathbb E_{\mathcal U}[p]
\ge
1-\eta-O(N^{-6})$ and
$\mathbb E_{\mathcal G}[p]
\le
-1+\eta+O(N^{-6}).$
Therefore,
\[
\left|
\mathbb E_{\mathcal U}[p]
-
\mathbb E_{\mathcal G}[p]
\right|
\ge
2(1-\eta)-O(N^{-6})
=
\Omega_\eta(1).
\]
Combining this with
\eqref{eq:forrelation-advantage-upper} yields
$\Omega_\eta(1)
\le
O({d^2}/{\sqrt N} )
+o(N^{-1/2}),$
and hence
$d=\Omega_\eta(N^{1/4}).$
\end{proof}

By Corollary \ref{cor:dqc1-sens-lower-bound}, we obtain the following result.

\begin{corollary}
\label{cor:two-forrelation-dqc1}
For every fixed $\eta\in(0,1/2)$, the
$2$-Forrelation promise above satisfies
\be
Q_\eta^{\mathrm{DQC1}}
(\mathrm{Forr}_{2,N})
=
\Omega_\eta(N^{1/4}).
\ee
\end{corollary}

The reduction from Fourier-growth to Forrelation used above is due to Girish, Raz, and Zhan~\cite{GirishRazZhan2021}. The point here is that the required level-two Fourier-growth bound follows directly from degree and coordinate sensitivity, so the lower bound can be recovered entirely within the sensitivity-degree framework.

\section{Conclusion}

This work shows that the polynomial method can be adapted to
the DQC1 query model by supplementing polynomial degree with a
coordinate sensitivity constraint, providing a tractable framework for
studying the limitations of DQC1 query algorithms.

Many questions remain open. 
%beyond those raised in
%Section~\ref{sec:symmetric-functions}.
% and~\ref{sec:sens-composition-lower}.
A fundamental goal is to obtain a more complete characterization of DQC1 query complexity. The sensitivity approximate degree introduced here captures only part of the structure satisfied by DQC1 output polynomials. As in the standard quantum query setting, we do not expect sensitivity approximate degree to characterize DQC1 query complexity in full generality. It is plausible that the construction of Ambainis and Belovs \cite{AmbainisBelovs2023} can be adapted to the DQC1 query model to yield an exponential separation between the two quantities.
So it is natural to ask what additional algebraic or matrix analytic constraints are needed for a more faithful characterization of DQC1 query algorithms. One possible direction is to seek a DQC1 analogue of the completely bounded forms that characterize standard quantum query algorithms \cite{arunachalam2019quantum}.

More broadly, one may seek stronger lower bound methods that exploit the tracial and mixed state structure of DQC1 beyond coordinate sensitivity, such as the adversary method or the dual polynomial method.
Such methods could potentially lead to sharper lower bounds for problems such as Forrelation and help clarify which quantum query advantages can survive in the one clean qubit setting.

% Finally, the folklore average-case classical simulation conjecture
% can be viewed as a structural statement about quantum query algorithms.
% Having established its DQC1 analogue, a natural direction for future
% research is to develop more refined structural theorems for DQC1
% query algorithms. In particular, we would like to understand what
% structural properties allow DQC1 algorithms to achieve exponential
% quantum-classical query separations, despite their strong
% average-case classical simulability.

\paragraph{Acknowledgment.}
The author would like to thank Ronald de Wolf for helpful comments on an earlier version of this paper, and Uma Girish for helpful correspondence and clarifications regarding the Fourier-growth bounds for DQC1 algorithms. This work was supported by the National Key Research Project of China under Grant No. 2025YFA1017200. 

\paragraph{AI disclosure.}
ChatGPT was used during the research leading to this paper. The main ideas and research directions were formulated by the author, while ChatGPT played a substantial role in developing proofs, carrying out calculations, and exploring extensions of known techniques from approximate degree and quantum query complexity.  Its use significantly accelerated the technical development of the work.  The author has reviewed and verified the arguments presented in the paper and takes full responsibility for their correctness.

\bibliographystyle{plain}
\bibliography{ref.bib}

\appendix

\end{document}